\documentclass[11pt,a4paper]{article}
\usepackage[utf8]{inputenc}
\usepackage[T1]{fontenc}
\usepackage{lmodern}
\usepackage[english]{babel}
\usepackage[margin=2.5cm]{geometry} 
\usepackage{graphicx} 
\usepackage{booktabs} 
\usepackage{natbib} 
\usepackage{amsmath} 
\usepackage{amssymb} 
\usepackage{setspace} 
\usepackage{caption} 
\usepackage[colorlinks=true, citecolor=blue, linkcolor=blue, urlcolor=blue]{hyperref} 

\RequirePackage{amsthm,amsfonts}
\RequirePackage[capitalize]{cleveref}
\RequirePackage{booktabs,multirow,makecell,tabularx}
\allowdisplaybreaks[4]

\theoremstyle{plain}
\newtheorem{theorem}{Theorem}
\newtheorem{lemma}{Lemma}
\newtheorem{corollary}{Corollary}
\theoremstyle{remark}

\newtheorem{example}{Example}
\newtheorem{condition}{Condition}

\def\v{{\varepsilon}}

\def\be{\begin{equation}}
\def\ee{\end{equation}}
\def\bea{\begin{eqnarray}}
\def\eea{\end{eqnarray}}
\def\pr{P}
\DeclareMathOperator{\var}{var}
\DeclareMathOperator{\cov}{cov}
\DeclareMathOperator{\tr}{tr}
\DeclareMathOperator{\ub}{ub}
\DeclareMathOperator{\amse}{amse}
\DeclareMathOperator{\diag}{diag}
\DeclareMathOperator{\skeww}{skew}
\DeclareMathOperator{\kurt}{kurt}
\DeclareMathOperator{\gaus}{gaus}
\DeclareMathOperator{\nong}{nong}
\DeclareMathOperator{\vecc}{vec}
\DeclareMathOperator{\inv}{inv}

\newcolumntype{Y}{>{\raggedleft\arraybackslash}X}

\setcitestyle{round}

\let\oldfootnote\footnote
\renewcommand{\footnote}{\fontsize{9}{11}\selectfont\oldfootnote}

\title{Errors-in-variables regression with estimated cross-observation covariance matrix: To prewhiten or not?}
\author{
    Jingkun Qiu \\
    Guanghua School of Management, Peking University, Beijing, China \\
    \and
    Hanyue Chen \\
    Center for Statistical Science, Peking University, Beijing, China \\
    \and
    Song Xi Chen\thanks{Corresponding author: sxchen@tsinghua.edu.cn} \\
    Department of Statistics and Data Science, Tsinghua University, Beijing, China
}
\date{} 

\begin{document}

\maketitle

\begin{abstract}
We consider statistical inference for errors-in-variables regression models with dependent observations under the high dimensionality of the error covariance matrix.
It is tempting to prewhiten the model and data that had led to efficient weighted least squares estimation in the presence of the measurement errors, as being practised in the optimal fingerprinting approach in climate change studies.
However, it is unclear to what extent the prewhitened estimator can improve the estimation efficiency of the unprewhitened estimator for errors-in-variables regression.
We compare the prewhitening and unprewhitening estimators in terms of their estimation efficiency and computational cost.
It shows that while the prewhitening operation does not necessarily improve the estimation efficiency of its unprewhitening counterpart, it demands more on the ensemble size needed in the error-covariance matrix estimation to ensure the asymptotic normality, and hence it would requires much more computationally resource.
\end{abstract}




\section{Introduction}
Errors-in-variables regression models 
have been widely used in modern scientific research from geophysical surveying to clinical data analysis, in which the regression covariates cannot be directly observed and only their imperfectly measured versions are available.
\cite{Gle81-MR0600530} established the asymptotic properties  for the estimator of 
regression coefficients, known as the total least squares or orthogonal regression estimator, for independent observations. 
We refer to \cite{Ful87-MR898653} for a comprehensive exposition of the related theory. 
However, a sizeable number of real world applications involve temporal and/or spatial dependence among observations, which makes the classical results for independent data unsuitable for direct application.

Climate change studies provide a   
different landscape, where the observations typically admit spatio-temporal dependence and the error covariance matrix is far more complex.
Motivated by the good fortune of the generalized least squares in the absence of the measurement errors, climate scientists are determined to prewhiten the errors-in-variables models with an estimated {error} covariance matrix, based on an ensemble of the random errors generated from the pre-industrial control simulations \citep{EBM16}.
This led to the optimal fingerprinting approach for detecting and attributing climate changes as advocated in \cite{AT99} and \cite{AS03}.
The method has been adopted as a technical foundation in the influential assessment reports by the Intergovernmental Panel on Climate Change \citep{IPCC6}.

The dependent observations and the measurement errors bring two challenging aspects. One is that, since the {measurement errors} lead us to a regime no longer blessed by the Gauss--Markov Theorem, do we still need to do the prewhitening as in the usual generalized least squares?
{To the best of our knowledge,} it is unclear whether or not and to what extent the prewhitened estimator can improve the statistical efficiency of the organic unprewhitened estimator. 
The other is that the sample size of dependent observations is the same as the dimension of the error covariance matrix, which implies that the error covariance matrix is automatically high dimensional and can have a rather complex structure.
The latter is different from the conventional errors-in-variables framework of \cite{Gle81-MR0600530} and \cite{Ful87-MR898653}, where the error covariance matrix admits simple diagonal structure.
{Note that the dimension of the error covariance matrix should not be confused with the dimension of covariates, which is typically fixed and not large  in climate change studies, representing anthropogenic (green house gases and aerosols) and natural factors, respectively.}

In the prewhitening operation, the climate scientists estimate the error covariance matrix by generating ensembles from the pre-industrial control simulations that represent a stable quasi-equilibrium climate state around the year 1850.
However, as the climate models are exceedingly expensive to run, the number of ensemble members is generally much smaller than the dimension of the covariance matrix, which further aggravates the high dimensional issue in the estimation of the error covariance matrix.

In this paper, we derive the asymptotic theory for the unprewhitened and the prewhitened estimators for the errors-in-variables model and provide a comprehensive comparison between the two estimators.
First, we extend the independent and identically distributed results in \cite{Gle81-MR0600530} for dependent data in Section \ref{sec:main} so that the asymptotic properties of the two estimators can be derived (Theorems \ref{thm:TLS-estimator} and \ref{thm:WTLS-estimator}).
In particular, to establish the asymptotic theory of the prewhitened estimator, a new perturbation bound for the generic weighted estimator with respect to the weighting matrix is derived (Theorem \ref{thm:PerturbationOfTheTLSEstimator}).
Second, we evaluate both estimators in terms of their statistical efficiencies (Section \ref{sec:stat-efficiency}) and computational costs (Section \ref{sec:comp-cost}).
It turns out that the optimal weighting matrix is not necessarily the one used in both estimators (Example \ref{ex:counterexample-1}) and that the two estimators have comparable asymptotic efficiency properties as neither of them can be more asymptotically efficient than the other (Examples \ref{cex:beat} and \ref{cex:beat-2}).
However, the unprewhitened estimator works more robustly than the prewhitened estimator when there are few replicate errors available due to limited computational resources.
Simulation studies are conducted in Section \ref{sec:simu} to demonstrate the aforementioned findings.
The proofs of Theorems \ref{thm:TLS-estimator}--\ref{thm:WTLS-estimator} are all deferred to Appendices \ref{pf:TLS}--\ref{pf:WTLS}.

\section{Main results}
\label{sec:main}
We begin with a brief introduction of useful notations.
Let $A\otimes B=(a_{ij}B)$ denotes the Kronecker product of $A$ and $B$ and $A\circ B=(a_{ij}b_{ij})$ denotes the Hadamard product.
Let $\lambda_{\min}(A)$ and $\lambda_{\max}(A)$ be, respectively, the smallest and the largest eigen values of $A$.
Let $\|x\|$ be the Euclidean norm of a vector $x$ and $\|A\|$ be the spectral norm of a matrix $A$.

Assume that the $p\times(m+1)$ data matrix $W=(z_{1},\dots,z_{m},y)$ is observed, where the response vector $y$ is of dimension $p\geq1$ that also represents the observation sample size, the covariate vectors $z_{1},\dots,z_{m}$ are, respectively, subject to measurement errors $u_{1},\dots,u_{m}$, and $p$ can grow to the infinity while $m$ is fixed.
Consider the model
\begin{align}
&y=\beta_{1}x_{1}+\dots+\beta_{m}x_{m}+\varepsilon,\label{eq:MER-y}\\&z_{j}=x_{j}+u_{j}\quad(j=1,\dots,m),\label{eq:MER-z}
\end{align}
where $x_{1},\dots,x_{m}$ are unobservable precise covariate vectors, $\varepsilon$ is the observation error vector, and the first two moments of the $p\times(m+1)$ augmented error matrix $\mathcal{E}=(u_{1},\dots,u_{m},\varepsilon)$ satisfy
\begin{equation}
E(\mathcal{E})=0\quad\text{and}\quad\var(\vecc\mathcal{E})=\begin{pmatrix}\Lambda\\
 & 1
\end{pmatrix}\otimes\Sigma.
\label{eq:MER-e}
\end{equation}
Let $\Psi=\diag(\Lambda,1)$.
Here, $\vecc\mathcal{E}=(u^{T}_{1},\dots,u^{T}_{m},\varepsilon^{T})^{T}$ denotes the vectorization of $\mathcal{E}$ in column-major order, the $p\times p$ cross-observation covariance matrix $\Sigma=\var(\varepsilon)$ is positive definite and accounts for dependence across different data points, and the $m\times m$ error variance ratio matrix $\Lambda=\var((u_{1i},\dots,u_{mi})^{T})/\var(\v_{i})$ is positive semidefinite and quantifies dependence across different measurement errors relative to the observation error for any single observation.
By semidefiniteness, we allow for the possibility of $\Lambda=0$ to accommodate the classical generalized least squares regression without measurement errors.

Throughout this paper, the cross-observation covariance matrix $\Sigma$ is estimated by $\hat{\Sigma}$ using an ensemble of replicate errors $v_{1},\dots,v_{n}\sim_{i.i.d.}\varepsilon$ where $n$ denotes the ensemble sample size, for instance the number of simulated runs from climate models in climate change studies, while the error variance ratio matrix $\Lambda$ is assumed to be known following the convention in the literature \citep{Kel84-MR733501, Ful87-MR898653}.
The justifications of the assumption for a known $\Lambda$ up to a scaling factor in geophysical surveys and psychometric testings have been discussed in \cite{Gle81-MR0600530}.
An estimator of $\Lambda$ in the presence of replicate measurements can be found in \cite{AF84-MR740908} and \cite{CRS95-MR1630517}.
In typical climate change studies, it is known that $\Lambda=\diag(1/n_{1},\dots,1/n_{m})$, where $n_{j}$ is the ensemble sample size for generating the $j$-th covariate vector $z_{j}$ as an ensemble mean \citep{AT99,AS03,LCY21}.

Let $\beta=(\beta_{1},\dots,\beta_{m})^{T}$, $Z=(z_{1},\dots,z_{m})$, and $X=(x_{1},\dots,x_{m})$ for compactness.
To encompass the prewhitening or otherwise, we define a weighted estimator for $\beta$ with a generic $p\times p$ symmetric weighting matrix $A$ as
\begin{equation}
\label{eq:TLS-estimator-general}
\hat{\beta}(A)=\hat{S}^{-1}_{xx}(A)S_{zy}(A),
\end{equation}
where $S_{zy}(A)=Z^{T}Ay/p$, and
\begin{equation}
\hat{S}_{xx}(A)=S_{zz}(A)-\frac{1}{\lambda_{\max}(S^{-1}_{ww}(A)\Psi)}\Lambda
\end{equation}
is a proxy of $S_{xx}(A)=X^{T}AX/p$ with $S_{zz}(A)=Z^{T}AZ/p$ and $S_{ww}({A})=W^{T}AW/p$.
Recall that $\hat{\Sigma}$ is an estimator of the cross-observation covariance matrix $\Sigma$.
Then, the unprewhitened and the prewhitened estimators are $\hat{\beta}(I_{p})$ and $\hat{\beta}(\hat{\Sigma}^{-1})$, respectively.
Note that $\hat{\beta}(I_{p})$ is the classical estimator used for independent and identically distributed observations \citep{Gle81-MR0600530, Ful87-MR898653} and that $\hat{\beta}(\hat{\Sigma}^{-1})$ is the prewhitened version of $\hat{\beta}(I_{p})$ via the transformation $W\mapsto\hat{\Sigma}^{-1/2}W$.

We first derive the asymptotic properties of the unprewhitened estimator $\hat{\beta}(I_{p})$.
Write $X=(X_{1},\dots,X_{p})^{T}$ and $\mathcal{E}=(\mathcal{E}_{1},\dots,\mathcal{E}_{p})^{T}$.
To account for the dependence among observations, we assume that the collection of the true covariates and the random errors, namely $\{(X_{i}^{T},\mathcal{E}_{i}^{T})^{T}:i\geq1\}$, admits a representation as a random field indexed by $D\subset\mathbb{R}^{d}$ for $d\geq1$ fixed.
Let $\{\xi_{s}:s\in D\}$ be the represented random field such that $\xi_{s}=(X_{i}^{T},\mathcal{E}_{i}^{T})^{T}$ if and only if $s=\varsigma(i)$ for a bijection $\varsigma:\mathbb{N}_{>0}\to D$.
For $I,J\subset D$, let $\rho(I,J)=\inf\{\|s-t\|:s\in I,t\in J\}$, and let $\sigma(I)$ be the $\sigma$-algebra generated by the $\{\xi_{s}:s\in I\}$.
Following \cite{Ros56-MR74711}, \cite{Dou94-MR1312160}, and \cite{JP09-MR2525996}, we define the $\alpha$-mixing coefficients $\alpha(k)=\sup_{\rho(I,J)\geq k}\sup_{A\in\sigma_{I},B\in\sigma_{J}}|\pr(A\cap B)-\pr(A)\pr(B)|$ for $k\geq1$.
Then, we impose the following regularity conditions.

\begin{condition}
\label{asp:moment}
(i) The collections $\{\|X_{i}\|^{4+2\delta}:i\geq1\}$ and $\{\|\mathcal{E}_{i}\|^{4+2\delta}:i\geq1\}$ are uniformly integrable for a constant $0<\delta<1$; and (ii) the first four moments of $\vecc\mathcal{E}$, conditioning on $X$, are the same as those of $N(0,\Psi\otimes\Sigma)$.
\end{condition}

\begin{condition}
\label{asp:mixing}
(i) The minimum distance of $D$ satisfies $\inf\{\|s-t\|:s,t\in D\}\geq\rho_{0}$ for a constant $\rho_{0}>0$; and (ii) the $\alpha$-mixing coefficients satisfy $\sum_{k=1}^{\infty}k^{(1+2/\delta)d-1}\alpha(k)<\infty$ for the same constant $\delta$ appeared in Condition \ref{asp:moment}(i).
\end{condition}

\begin{condition}
\label{asp:trace}
(i) The moment limits of $E\{S_{xx}(I_{p})\}$ and $E\{S_{xx}(\Sigma)\}$ exist as $p\to\infty$ and are positive definite matrices, denoted by $ Q_{0}$ and $ Q_{1}$ respectively; and
(ii) The trace limits of $\tr(\Sigma)/p$ and $\tr(\Sigma^{2})/p$ exist as $p\to\infty$ and are positive, denoted by $\tau_{1}$ and $\tau_{2}$ respectively.
\end{condition}

Conditions \ref{asp:moment}(i) and \ref{asp:mixing} are required to apply the law of large numbers and the central limit theorem for random fields \citep{JP09-MR2525996}.
Condition \ref{asp:trace} is a direct generalization for dependent observations of that in \cite{Gle81-MR0600530}.
As in \cite{Gle81-MR0600530} for independent observations, we require the Gaussian moment matching condition \ref{asp:moment}(ii) to derive an explicit expression of the asymptotic covariance matrix of the unprewhitened estimator for a clear comparison with that of the prewhitened estimator.
This condition will be relaxed in Section \ref{sec:non-Gaussian} with non-Gaussian higher order moments.

\begin{theorem}
\label{thm:TLS-estimator}
(i) Under Conditions \ref{asp:moment}--\ref{asp:trace} for Model \eqref{eq:MER-y}--\eqref{eq:MER-e}, $S_{ww}(I_{p})$ and $\hat{S}_{xx}(I_{p})$ are nonsingular and thus $\hat{\beta}(I_{p})$ is well defined with probability approaching to one as $p\to\infty$.
Furthermore,
\begin{equation}
\label{eq:TLS-estimator-asymptotic-normality}
\hat{\beta}(I_{p})\overset{\pr}{\to}\beta\quad\text{and}\quad p^{1/2}\{\hat{\beta}(I_{p})-\beta\}\overset{d}{\to}N(0,\Omega), 
\end{equation}
where
\begin{equation}
\label{eq:Omega}
\Omega=(1+\beta^{T}\Lambda\beta)Q^{-1}_{0}\Big\{ Q_{1}+\tau_{2}\Big(\Lambda-\frac{\Lambda\beta\beta^{T}\Lambda}{1+\beta^{T}\Lambda\beta}\Big)\Big\} Q^{-1}_{0}.
\end{equation}
In particular, we have $\hat{S}_{xx}(I_{p})\to_{\pr}Q_{0}$.

(ii) Let $\hat{\Sigma}$ be an estimator of $\Sigma$ such that $\|\hat{\Sigma}-\Sigma\|=o_{\pr}(1)$.
If Conditions \ref{asp:moment}--\ref{asp:mixing} are valid with $W$ replaced by $\Sigma^{1/2}W$, then we have $\hat{S}_{xx}(\hat{\Sigma})\to_{\pr}Q_{1}$ and $\hat{\tau}_{2}:=\tr(\hat{\Sigma}^{2})/p\to_{\pr}\tau_{2}$, respectively.
\end{theorem}

Note that the consistent estimators of $Q_{0}$, $Q_{1}$, and $\tau_{2}$ in Theorem \ref{thm:TLS-estimator} can be used to provide a plug-in and  also consistent estimator of $\Omega$, given by
\begin{equation}
\hat{\Omega}=\{1+\hat{\beta}^{T}(I_{p})\Lambda\hat{\beta}(I_{p})\}\hat{S}^{-1}_{xx}(I_{p})\Big[\hat{S}_{xx}(\hat{\Sigma})+\hat{\tau}_{2}\Big\{\Lambda-\frac{\Lambda\hat{\beta}(I_{p})\hat{\beta}^{T}(I_{p})\Lambda}{1+\hat{\beta}^{T}(I_{p})\Lambda\hat{\beta}(I_{p})}\Big\}\Big]\hat{S}^{-1}_{xx}(I_{p}).
\end{equation}
{Then}, a confidence region based on the unprewhitened estimator $\hat{\beta}(I_{p})$ {can be constructed}.
When the weighting matrix $A$ in \eqref{eq:TLS-estimator-general} is nonrandom, the asymptotic properties of the general weighted estimator $\hat{\beta}(A)$ are direct consequences of those in Theorem \ref{thm:TLS-estimator}.
{The following Condition \ref{asp:trace-WTLS} is a direct generalization of Condition \ref{asp:trace} for a generic symmetric $A$.}

\begin{condition}
\label{asp:trace-WTLS}
(i) The moment limits of $E\{S_{xx}(A)\}$ and $E\{S_{xx}(A\Sigma A)\}$ exist as $p\to\infty$ and are positive definite matrices, denoted by $ {Q}_{0}^{*}$ and $ {Q}_{1}^{*}$ respectively; and
(ii) The trace limits of $\tr(\Sigma A)/p$ and $\tr (\Sigma A\Sigma A)/p$ exist as $p\to\infty$ and are positive, denoted by $\tau_{1}^{*}$ and $\tau_{2}^{*}$ respectively.
\end{condition}

\begin{corollary}
\label{cor:TLS-estimator-general}
Under Conditions \ref{asp:moment}--\ref{asp:mixing} (with $W$ replaced by $A^{1/2}W$) and \ref{asp:trace-WTLS} for Model \eqref{eq:MER-y}--\eqref{eq:MER-e}, $S_{ww}(A)$ and $\hat{S}_{xx}(A)$ are nonsingular and thus $\hat{\beta}(A)$ is well defined with probability approaching to one as $p\to\infty$.
Furthermore,
\begin{equation}
\label{eq:WTLS-estimator-asymptotic-normality-auxiliary}
\hat{\beta}(A)\overset{\pr}{\to}\beta\quad\text{and}\quad p^{1/2}\{\hat{\beta}(A)-\beta\}\overset{d}{\to}N(0,\Omega^{*}),
\end{equation}
where
\begin{equation*}
\Omega^{*}=(1+\beta^{T}\Lambda\beta)(Q_{0}^{*})^{-1}\Big\{Q_{1}^{*}+\tau_{2}^{*}\Big(\Lambda-\frac{\Lambda\beta\beta^{T}\Lambda}{1+\beta^{T}\Lambda\beta}\Big)\Big\}(Q_{0}^{*})^{-1}.
\end{equation*}
In particular, if $A=\Sigma^{-1}$ and $E\{S_{xx}(\Sigma^{-1})\}\to Q_{0}^{**}$ that is positive definite, then $\Omega^{*}$ becomes
\begin{equation}
\Omega^{**}=(1+\beta^{T}\Lambda\beta)(Q_{0}^{**})^{-1}\Big(Q_{0}^{**}+\Lambda-\frac{\Lambda\beta\beta^{T}\Lambda}{1+\beta^{T}\Lambda\beta}\Big)(Q_{0}^{**})^{-1}.
\end{equation}
\end{corollary}

To analyze the prewhitened estimator $\hat{\beta}(\hat{\Sigma}^{-1})$, we observe that
\begin{equation}
\label{eq:TLS-decomposition}
\hat{\beta}(\hat{\Sigma}^{-1})-\beta=\hat{\beta}(\hat{\Sigma}^{-1})-\hat{\beta}(\Sigma^{-1})+\hat{\beta}(\Sigma^{-1})-\beta.
\end{equation}
The asymptotic properties of the second part $\hat{\beta}(\Sigma^{-1})-\beta$ are immediate consequences of Corollary \ref{cor:TLS-estimator-general} with the case of $A=\Sigma^{-1}$.
Then, it suffices to investigate the first part $\hat{\beta}(\hat{\Sigma}^{-1})-\hat{\beta}(\Sigma^{-1})$ of \eqref{eq:TLS-decomposition}.
To this end, we shall establish a perturbation bound for the weighted estimator $\hat{\beta}(A)$ with respect to the weighting matrix $A$.

Recall that $S_{ww}(A)=W^{T}AW/p$, $S_{zz}(A)=Z^{T}AZ/p$, $S_{zy}(A)=Z^{T}Ay/p$, and $S_{yy}(A)=y^{T}Ay/p$, $\hat{\lambda}(A)=1/\lambda_{\max}(S_{ww}^{-1}(A)\Psi)$, and $\hat{S}_{xx}(A)=S_{zz}(A)-\hat{\lambda}(A)\Lambda$.
Let $\Delta(B,A)=\|A^{-1}\|\|B-A\|$.
We notice that the following two quantities are important in the perturbation bound:
\begin{equation}
\ub(\Delta(\hat{S}_{xx}(B),\hat{S}_{xx}(A)))=\frac{\lambda_{\max}(S_{zz}(A))+\hat{\lambda}(A)\lambda_{\max}(\Lambda)}{\lambda_{\min}(\hat{S}_{xx}(A))}\Delta(B,A),
\end{equation}
and
\begin{equation}
\ub(\|\hat{\beta}(A)\|)=\frac{\lambda^{1/2}_{\max}(S_{zz}(A))S^{1/2}_{yy}(A)}{\lambda_{\min}(\hat{S}_{xx}(A))}.
\end{equation}
Observe that $\ub(\|\hat{\beta}(A)\|)$ serves as an upper bound of $\|\hat{\beta}(A)\|$.
As will be shown in the proof of the following theorem, $\ub(\Delta(\hat{S}_{xx}(B),\hat{S}_{xx}(A)))$ is also an upper bound of $\Delta(\hat{S}_{xx}(B),\hat{S}_{xx}(A))$, which measures the distance between $\hat{S}_{xx}(B)$ and $\hat{S}_{xx}(A)$, and is the driving factor of the distance between $\hat{\beta}(B)$ and $\hat{\beta}(A)$.
Clearly, we have $\Delta(B,A)\leq\ub(\Delta(\hat{S}_{xx}(B),\hat{S}_{xx}(A)))$.
The next theorem provides a perturbation bound for $A\mapsto\hat{\beta}(A)$ over real symmetric matrices.

\begin{theorem}
\label{thm:PerturbationOfTheTLSEstimator}
Let $A$ and $B$ be two $p\times p$ real symmetric matrices.
Assume that $A$, $S_{ww}(A)$, and $\hat{S}_{xx}(A)$ are positive definite (so that $\hat{\beta}(A)$ is well defined), and $\hat{\Delta}=\ub(\Delta(\hat{S}_{xx}(B),\hat{S}_{xx}(A)))<1$.
Then, $B$, $S_{ww}(B)$, and $\hat{S}_{xx}(B)$ are also positive definite (so that $\hat{\beta}(B)$ is also well defined).
Furthermore,
\begin{equation}
\label{eq:thm-PerturbationOfTheTLSEstimator}
\|\hat{\beta}(B)\|\leq\frac{1+\hat{\Delta}}{1-\hat{\Delta}}\ub(\|\hat{\beta}(A)\|)\quad\text{and}\quad
\|\hat{\beta}(B)-\hat{\beta}(A)\|\leq\frac{2\hat{\Delta}}{1-\hat{\Delta}}\ub(\|\hat{\beta}(A)\|).
\end{equation}
\end{theorem}

Under the conditions of Corollary \ref{cor:TLS-estimator-general} with $A=\Sigma^{-1}$, it follows from the corollary that $\hat{\beta}(\Sigma^{-1})$ is well defined with probability approaching to one.
When $\|\hat{\Sigma}^{-1}-\Sigma^{-1}\|$ is small, it can be expected that so is $\ub(\Delta(\hat{S}_{xx}(\hat{\Sigma}^{-1}),\hat{S}_{xx}(\Sigma^{-1})))$ and thus by Theorem \ref{thm:PerturbationOfTheTLSEstimator}, the prewhitened estimator $\hat{\beta}(\hat\Sigma^{-1})$ is also well defined with probability approaching to one.
Indeed, the asymptotic properties of $\hat{\beta}(\hat\Sigma^{-1})$ can be provided as follows.

\begin{theorem}
\label{thm:WTLS-estimator}
Let $\hat{\Sigma}^{-1}$ be an estimator of $\Sigma^{-1}$ such that $\|\Sigma\|\|\hat{\Sigma}^{-1}-\Sigma^{-1}\|=o_{\pr}(p^{-1/2})$.
Under the conditions of Corollary \ref{cor:TLS-estimator-general} with $A=\Sigma^{-1}$ for Model \eqref{eq:MER-y}--\eqref{eq:MER-e}, $S_{ww}(\hat\Sigma^{-1})$ and $\hat{S}_{xx}(\hat\Sigma^{-1})$ are nonsingular and thus $\hat{\beta}(\hat\Sigma^{-1})$ is well defined with probability approaching to one as $p\to\infty$.
Furthermore,
\begin{equation}
\label{eq:WTLS-estimator-asymptotic-normality}
\hat{\beta}(\hat{\Sigma}^{-1})\overset{\pr}{\to}\beta\quad\text{and}\quad p^{1/2}\{\hat{\beta}(\hat{\Sigma}^{-1})-\beta\}\overset{d}{\to}N(0,\Omega^{**}),
\end{equation}
where $\Omega^{**}$ is defined in Corollary \ref{cor:TLS-estimator-general}.
In particular, we have $\hat{S}_{xx}(\hat{\Sigma}^{-1})\to_{\pr}Q_{0}^{**}$.
\end{theorem}

We emphasize that the estimated cross-observation covariance matrix $\hat\Sigma$ does not need to be independent of the observed data matrix $W$.
The only requirement for $\hat\Sigma$ in Theorem \ref{thm:WTLS-estimator} is to be a consistent estimator of $\Sigma$ with a fast enough rate of convergence.
This provides an additional flexibility to use a broader class of ensembles of replicate errors $v_{1},\dots,v_{n}$ used to generate $\hat\Sigma$ in practice.
Indeed, as documented in \citet[pages 1951--1952]{EBM16}, it is recommended to construct the pre-industrial control simulations (namely the replicate errors in climate change studies) using a background volcanic aerosol (e.g., 1850--2014 mean) to avoid artefacts and artificial transient effects in simulations, which may result in a possible statistical dependency between $\hat\Sigma$ and $W$ through the information leakage.

\subsection{Evaluation via statistical efficiency}
\label{sec:stat-efficiency}
We evaluate the statistical efficiencies of $\hat{\beta}(I_{p})$ and $\hat{\beta}(\hat{\Sigma}^{-1})$.
Under the Gauss--Markov theorem,  the optimal weighting matrix for the linear regression problem without measurement errors is $c\Sigma^{-1}$ for an arbitrary constant $c>0$.
For the errors-in-variables model \eqref{eq:MER-y}--\eqref{eq:MER-e}, a natural conjecture is that $\hat{\beta}(\hat{\Sigma}^{-1})$ were the efficient estimator.
Indeed, this has been widely and implicitly conjectured in the climate change community \citep{AS03, RPT13, DTY19},
where the estimator led to the ``optimal'' fingerprinting approach.

The next counterexample shows that the conjecture is false as the optimal weighting matrix $A$ for $\hat{\beta}(A)$ is neither $cI_{p}$ nor $c\Sigma^{-1}$ in general.
For ease of exposition, we shall only consider the case of $m=1$ and $\Lambda=I_{m}=1$ with the independent but heteroscedastic observations such that $\Sigma$ is diagonal. 
Recall that the asymptotic mean squared error of an estimator $T_{n}$ for a parameter $\theta$ is $\amse(T_{n})=E(T^{2})/a_{n}^{2}$ when $a_{n}(T_{n}-\theta)\to_{d}T$ as $n\to\infty$ and a positive sequence $\{a_{n}\}$ \citep[page 138]{Sha03-MR2002723}.

\begin{example} 
\label{ex:counterexample-1}
Let $m=1$, $\beta_{1}\in\mathbb{R}$, $x_{1}=(x_{11},\dots,x_{1p})^{T}$ for $x_{1i}\neq0$, $\Sigma=\diag(\sigma_{1}^{2},\dots,\sigma_{p}^{2})$ for $\sigma_{i}^{2}>0$, $A=\diag(a_{1},\dots,a_{p})$ for $a_{i}>0$, and $A^{*}=\diag(a_{1}^{*},\dots,a_{p}^{*})$ for $a_{i}^{*}=x_{1i}^{2}/\{(1+\beta_{1}^{2})x_{1i}^{2}\sigma_{i}^{2}+\sigma_{i}^{4}\}$ ($i=1,\dots,p$).
By Corollary \ref{cor:TLS-estimator-general},
\begin{align}
\amse(\hat{\beta}(A))&=\frac{\sum_{i=1}^{p}a_{i}^{2}\{(1+\beta_{1}^{2})x_{1i}^{2}\sigma_{i}^{2}+\sigma_{i}^{4}\}}{(\sum_{i=1}^{p}a_{i}x_{1i}^{2})^{2}}\notag\\&\geq\frac{1}{\sum_{i=1}^{p}x_{1i}^{4}/\{(1+\beta_{1}^{2})x_{1i}^{2}\sigma_{i}^{2}+\sigma_{i}^{4}\}}=\amse(\hat{\beta}(A^{*})),
\label{eq:counterexample-1}
\end{align}
where the equality holds if and only if $A=cA^{*}$ for an arbitrary constant $c>0$.
This implies that $A^{*}$ is the optimal diagonal weighting matrix and that $A^{*}\neq cI_{p}$ and $A^{*}\neq c\Sigma^{-1}$ in general.
\end{example}

As the optimal weighting matrix $A^{*}$ depends on the knowledge of $\beta_{1}$, $x_{1}$, and $\Sigma$,
the suboptimal choices of $\hat{\beta}(I_{p})$ and $\hat{\beta}(\hat{\Sigma}^{-1})$ are still useful in the real world application.
In particular, both estimators can respectively match $\hat{\beta}(A^{*})$ in Example \ref{ex:counterexample-1} for particular specifications of $x_{1}$.
Indeed, we have $A^{*}=cI_{p}$ if and only if $x_{1i}^{2}=\sigma_{i}^{4}/\{c-(1+\beta_{1}^{2})\sigma_{i}^{2}\}$ ($i=1,\dots,p$) while $A^{*}=c\Sigma^{-1}$ if and only if $x_{1i}^{2}=c\sigma_{i}^{2}$ ($i=1,\dots,p$).

However, the efficiency order of the two estimators is much more ambiguous.
The following two examples further demonstrate that neither of them can be more asymptotically efficient than the other unless the relation between $\beta_{1}$, $x_{1}$, and $\Sigma$ is known and carefully specified.
In contrast to the Gauss--Markov setting, the statistical efficiency gain of the prewhitening transformation $W\mapsto\hat{\Sigma}^{-1/2}W$ in the measurement error problem cannot always be ensured.

\begin{example}[$\hat{\beta}(I_{p})$ can be better than $\hat{\beta}(\hat{\Sigma}^{-1})$]
\label{cex:beat}
Let $m=1$, $\beta_{1}\in\mathbb{R}$, $x_{1}=(x_{11},\dots,x_{1p})^{T}$, and $\Sigma=\diag(\sigma_{1}^{2},\dots,\sigma_{p}^{2})$ for $0<\sigma_{i}^{2}<\sigma_{\max}^{2}$, $x_{1i}^{2}=\sigma_{i}^{4}/\{(1+\beta_{1}^{2})(\sigma_{\max}^{2}-\sigma_{i}^{2})\}$, $c_{i}=x_{1i}^{2}/\sigma_{i}^{2}$ ($i=1,\dots,p$), and $\sigma_{\max}^{2}>0$.
By Theorems \ref{thm:TLS-estimator} and \ref{thm:WTLS-estimator},
\begin{align}
\amse(\hat{\beta}(\hat{\Sigma}^{-1}))&=\frac{\sum_{i=1}^{p}\{1+(1+\beta_{1}^{2})c_{i}\}}{(\sum_{i=1}^{p}c_{i})^{2}}\notag\\&\geq\frac{1}{\sum_{i=1}^{p}c_{i}^{2}/\{1+(1+\beta_{1}^{2})c_{i}\}}=\amse(\hat{\beta}(I_{p})),
\label{eq:counterexample-3}
\end{align}
where the equality holds if and only if $\Sigma=\sigma^{2}I_{p}$ for an arbitrary constant $0<\sigma^{2}<\sigma_{\max}^{2}$.
\end{example}

\begin{example}[$\hat{\beta}(\hat{\Sigma}^{-1})$ can be better than $\hat{\beta}(I_{p})$]
\label{cex:beat-2}
Let $m=1$, $\beta_{1}\in\mathbb{R}$, $x_{1}=(x_{11},\dots,x_{1p})^{T}$, and $\Sigma=\diag(\sigma_{1}^{2},\dots,\sigma_{p}^{2})$ for $x_{1i}^{2}=\sigma_{i}^{2}>0$ ($i=1,\dots,p$).
Then
\begin{equation}
\amse(\hat{\beta}(\hat{\Sigma}^{-1}))=\frac{2+\beta_{1}^{2}}{p}\leq\frac{(2+\beta_{1}^{2})\sum_{i=1}^{p}\sigma_{i}^{4}}{(\sum_{i=1}^{p}\sigma_{i}^{2})^{2}}=\amse(\hat{\beta}(I_{p})),
\label{eq:counterexample-4}
\end{equation}
where the equality holds if and only if $\Sigma=\sigma^{2}I_{p}$ for an arbitrary constant $\sigma^{2}>0$.
\end{example}

\subsection{Evaluation via computational cost}
\label{sec:comp-cost}
Recall that the estimator $\hat{\Sigma}$ of the error covariance matrix $\Sigma$ is obtained from the replicate errors $v_{1},\dots,v_{n}\sim_{i.i.d.}\varepsilon$ in the errors-in-variables model \eqref{eq:MER-y}--\eqref{eq:MER-e}, where the ensemble sample size $n$ is typically much smaller than the observation sample size $p$ due to the limited computational resources used to generate those ensembles in practice.
This encourages us to compare the computational costs of the two estimators, namely the requirements for the growth rate of $n$ in $p$ to ensure a valid statistical inference.
Assume that $1/c\leq\lambda_{\min}(\Sigma)\leq\lambda_{\max}(\Sigma)\leq c$ for a constant $c>0$.
In view of Theorems 1 and 3 in \cite{CZZ10-MR2676885}, the minimax optimal rate of convergence of $E(\|\hat{\Sigma}-\Sigma\|^{2})$ and $E(\|\hat{\Sigma}^{-1}-\Sigma^{-1}\|^{2})$ over a well-conditioned class is $\min\{n^{-2\alpha/(2\alpha+1)}+n^{-1}\log p,n^{-1}p\}$, where $\alpha>0$ quantifies the decaying rate of the off-diagonal elements of $\Sigma$.
By Theorems \ref{thm:TLS-estimator} and \ref{thm:WTLS-estimator}, the requirements for the rate of convergence of $\hat{\Sigma}$ to construct a valid confidence region based on $\hat{\beta}(I_{p})$ and $\hat{\beta}(\hat{\Sigma}^{-1})$ are $\|\hat{\Sigma}-\Sigma\|=o_{\pr}(1)
$ and $\|\hat{\Sigma}^{-1}-\Sigma^{-1}\|=o_{\pr}(p^{-1/2})$ respectively.
Then the (minimax) best possible ensemble sample size requirements of $\hat{\beta}(I_{p})$ and $\hat{\beta}(\hat{\Sigma}^{-1})$ are, respectively, $\log p=o(n)$ and $p^{\min\{2,(2\alpha+1)/2\alpha\}}=o(n)$.

As the required ensemble sample size $n$ of $\hat{\beta}(I_{p})$ is in a logarithmic rate of the observation sample size $p$ while that of $\hat{\beta}(\hat{\Sigma}^{-1})$ is in a polynomial rate, we can conclude that the prewhitened estimator is much more computationally expensive than the unprewhitened estimator.
Indeed, the requirement of $p^{\min\{2,(2\alpha+1)/2\alpha\}}=o(n)$ for $\hat{\beta}(\hat{\Sigma}^{-1})$ is an unrealistic condition as $n$ is typically much smaller than $p$ for most real world applications.
For instance, we note that $p=572$ and $n=223$ in the climate change study of \cite{LCY21}.
A direct application of the prewhitening transformation in this case may lead to unreliable empirical results, for instance an under-coverage confidence region.
Therefore, the unprewhitened estimator $\hat{\beta}(I_{p})$ is recommended instead.

\section{Extension for non-Gaussian higher order moments}
\label{sec:non-Gaussian}

The asymptotic normality of the unprewhitened estimator $\hat{\beta}(I_{p})$ established in Theorem \ref{thm:TLS-estimator} relies heavily on Condition \ref{asp:moment}(ii), namely the first four conditional moments of the augmented error matrix $\mathcal{E}$ match those of a corresponding Gaussian distribution.
While this condition leads to a simplified asymptotic covariance matrix $\Omega$ and a clear comparison between the unprewhitened and the prewhitened estimator as shown above, it can be too restrictive in practice.
In many real world applications, such as the temperature and the precipitation data in climate change studies, random errors may exhibit much more skewed and heavier tails than the Gaussian distribution.

To overcome the limitation of the Gaussian moment matching condition \ref{asp:moment}(ii), we observe from \eqref{eq:asymptotic-expression} in the proof of Theorem \ref{thm:TLS-estimator} that the $\hat{\beta}(I_{p})$ admits the following asymptotic expansion:
\begin{equation}
\label{eq:asymptotic-expansion}
\hat{\beta}(I_{p})-\beta=Q^{-1}_{0}(I_{m}+\Lambda\beta\beta^{T})^{-1}S_{g}\{1+o_{\pr}(1)\},
\end{equation}
where $S_{g}=p^{-1}\sum^{p}_{i=1}g_{i}$ and $g_{i}=(y_{i}-Z^{T}_{i}\beta)(Z_{i}+y_{i}\Lambda\beta)$.
Consequently, the asymptotic covariance matrix of $\hat{\beta}(I_{p})$ is determined by that of $S_{g}$, which is summarized in the following theorem.

\begin{condition}
\label{asp:moment-nonGaussian-general}
(i) The conditional moment constraints $E(\mathcal{E}\mid X)=0$ and $\var(\vecc\mathcal{E}\mid X)=\Psi\otimes\Sigma$ hold; and (ii) The moment limit of $\var(p^{1/2}S_{g})$ exist as $p\to\infty$, denoted by $V_{g}$.
\end{condition}

\begin{theorem}
\label{thm:TLS-estimator-nonGaussian-general}
Under Conditions \ref{asp:moment}(i), \ref{asp:mixing}--\ref{asp:trace}, and \ref{asp:moment-nonGaussian-general} for Model \eqref{eq:MER-y}--\eqref{eq:MER-e}, the results in Theorem \ref{thm:TLS-estimator} still hold except that the asymptotic covariance matrix $\Omega$ is replaced by
\begin{equation}
\label{eq:Omega-general}
\Omega_{\nong}=Q^{-1}_{0}(I_{m}+\Lambda\beta\beta^{T})^{-1}V_{g}(I_{m}+\beta\beta^{T}\Lambda)^{-1}Q^{-1}_{0}.
\end{equation}
\end{theorem}

Clearly, Theorem \ref{thm:TLS-estimator} is a special case of Theorem \ref{thm:TLS-estimator-nonGaussian-general}.
Indeed, Condition \ref{asp:moment-nonGaussian-general}(ii) can be deduced from Condition \ref{asp:moment}(ii), as under the latter condition,
\begin{equation}
\var(p^{1/2}S_{g})\to(1+\beta^{T}\Lambda\beta)\{(I_{m}+\Lambda\beta\beta^{T})Q_{1}(I_{m}+\beta\beta^{T}\Lambda)+\tau_{2}(\Lambda+\Lambda\beta\beta^{T}\Lambda)\},
\end{equation}
which is the desired moment limit $V_{g}$.
Plugging this limit back to \eqref{eq:Omega-general} recovers \eqref{eq:Omega}.

\subsection{Effects of non-Gaussianity through marginal moments}
Theorem \ref{thm:TLS-estimator-nonGaussian-general} allows us to investigate the effects of non-Gaussian third and fourth moments on the asymptotic covariance matrix of the unprewhitened estimator $\hat{\beta}(I_{p})$.
Denote $\diag(\Sigma)=(\sigma^{2}_{1},\dots,\sigma^{2}_{p})^{T}$ and $\diag(\Lambda)=(\lambda^{2}_{1},\dots,\lambda^{2}_{m})^{T}$.
The following two assumptions will be required to obtain a clear analysis.

\begin{condition}
\label{asp:moment-nonGaussian}
The first four moments of $\vecc\mathcal{E}$, conditioning on $X$, are the same as those of $N(0,\Psi\otimes\Sigma)$ except that $E(\mathcal{E}^{3}_{ij}\mid X)=\kappa_{3}\{E(\mathcal{E}^{2}_{ij}\mid X)\}^{3/2}$ and $E(\mathcal{E}^{4}_{ij}\mid X)=(3+\kappa_{4})\{E(\mathcal{E}^{2}_{ij}\mid X)\}^{2}$ for two constants $\kappa_3$ and $\kappa_4$.
\end{condition}

\begin{condition}
\label{asp:trace-more}
(i) The moment limit of $p^{-1}\sum^{p}_{i=1}\sigma^{3}_{i}E(X_{i})$ exists as $p\to\infty$, denoted by $\mu_{x}$; and
(ii) The trace limit of $\tr(\Sigma\circ\Sigma)/p$ exists as $p\to\infty$ and is positive, denoted by $\tau_{4}$.
\end{condition}

To model the departure from Gaussianity, instead of using Condition \ref{asp:moment-nonGaussian-general}, the two parameters $\kappa_3$ and $\kappa_4$ are introduced in Condition \ref{asp:moment-nonGaussian} to represent the skewness and the excess kurtosis of the random errors, respectively.
As noted by \cite{Kel84-MR733501} for independent observations, the asymptotic results can be involved under non-Gaussian cross-moments.
For ease of exposition, Condition \ref{asp:moment-nonGaussian} maintains the independence-like structure for the cross-moments up to the fourth order, but allows the marginal distributions of the random errors to possess a common skewness of $\kappa_3$ and an excess kurtosis of $\kappa_4$, where the Gaussian distribution corresponds to the special baseline case with $\kappa_3=0$ and $\kappa_4=0$.
To ensure that the contributions of these higher order moments stabilize asymptotically, Condition \ref{asp:trace-more} is additionally required, serving as a higher order moment analogue to Condition \ref{asp:trace}.
Under these assumptions, Theorem \ref{thm:TLS-estimator-nonGaussian-general} can be specialized as the following theorem.

\begin{theorem}
\label{thm:TLS-estimator-nonGaussian}
Under Conditions \ref{asp:moment}(i), \ref{asp:mixing}--\ref{asp:trace}, and \ref{asp:moment-nonGaussian}--\ref{asp:trace-more} for Model \eqref{eq:MER-y}--\eqref{eq:MER-e}, the results in Theorem \ref{thm:TLS-estimator} still hold except that the asymptotic covariance matrix $\Omega$ is replaced by $\Omega+\Omega_{\skeww}+\Omega_{\kurt}$, where
\begin{equation}
\Omega_{\skeww}=\kappa_{3}Q^{-1}_{0}(\mu_{x}\nu^{T}_{\beta}+\nu_{\beta}\mu^{T}_{x})Q^{-1}_{0},
\end{equation}
\begin{equation}
\Omega_{\kurt}=\kappa_{4}\tau_{4}Q^{-1}_{0}(I_{m}+\Lambda\beta\beta^{T})^{-1}(D_{\beta}+\Lambda\beta\beta^{T}\Lambda)(I_{m}+\beta\beta^{T}\Lambda)^{-1}Q^{-1}_{0},
\end{equation}
$\nu_{\beta}=(I_{m}+\Lambda\beta\beta^{T})^{-1}(\gamma_{\beta}+\Lambda\beta)$, $\gamma_{\beta}=(\lambda^{3}_{1}\beta^{2}_{1},\dots,\lambda^{3}_{m}\beta^{2}_{m})^{T}$, and $D_{\beta}=\diag(\lambda^{4}_{1}\beta^{2}_{1},\dots,\lambda^{4}_{m}\beta^{2}_{m})$.
\end{theorem}

Theorem \ref{thm:TLS-estimator-nonGaussian} decomposes the total asymptotic covariance matrix into three additive components, namely the baseline Gaussian-like component $\Omega$, a skewness-driven component $\Omega_{\skeww}$, and a kurtosis-driven component $\Omega_{\kurt}$.
When the random errors exhibit right skewed ($\kappa_3>0$) and leptokurtic ($\kappa_4>0$) tails and the true covariates have a positive weighted average with $\mu_x>0$, the both additional component $\Omega_{\skeww}$ and $\Omega_{\kurt}$ will be positive and thus the asymptotic covariance matrix of $\hat{\beta}(I_{p})$ can be severely inflated.
Consequently, the confidence region could be under-coverage in this case if the Gaussian-like region based on Theorem \ref{thm:TLS-estimator} were used.
Note also that the inflation effect can be alleviated if the true covariates are well centered such that $\mu_x=0$, resulting in a removal of $\Omega_{\skeww}$ in Theorem \ref{thm:TLS-estimator-nonGaussian}.

\subsection{Long run covariance matrix estimation}
\label{sec:hac}

While Theorem \ref{thm:TLS-estimator-nonGaussian} provides a comprehensive characterization of the asymptotic covariance matrix of $\hat{\beta}(I_{p})$ under non-Gaussian higher order moments, the estimation of $\kappa_3$, $\kappa_4$, and $\mu_x$ can be tedious in practice.
Furthermore, Condition \ref{asp:moment-nonGaussian} is still restrictive, as the cross-moments of random errors can also be non-Gaussian given the spatial dependency nature of the real world data.
This motivates the development of a robust covariance matrix estimation strategy that can automatically account for both the data dependency and the non-Gaussian distributions without requiring explicit estimation of any higher order moments.
The key is to directly estimate the long run covariance matrix $V_{g}$ in Theorem \ref{thm:TLS-estimator-nonGaussian-general}.
For the ordinary least squares regression with time series data, \cite{NW87-MR890864}, \cite{And91-MR1106513}, and \cite{NW94-MR1299308} proposed a set of consistent estimators of the $V_{g}$ type covariance matrices via the kernel estimation.
These results have been extended for spatially dependent data in \cite{Con99-MR1707000}, \cite{KP07-MR2395919}, and \cite{KS11-MR2748557}.

Recall that $s_i = \varsigma(i) \in D$ denote the spatial location of the $i$-th observation in Condition \ref{asp:mixing}.
In view of \eqref{eq:asymptotic-expansion}, a natural kernel estimator of $V_{g}$ is given by
\begin{equation}
\label{eq:HAC-estimator}
\hat{V}_{g} = \frac{1}{p} \sum_{i=1}^p \sum_{j=1}^p \hat{g}_i\hat{g}_j^T \mathcal{K}\Big(\frac{\|s_i - s_j\|}{h_p}\Big),
\end{equation}
where $\hat{g}_{i}=\{y_{i}-Z_{i}^{T}\hat{\beta}(I_{p})\}\{Z_{i}+y_{i}\Lambda\hat{\beta}(I_{p})\}$ is a natural proxy of $g_{i}$, $\mathcal{K}(\cdot)$ is a kernel function, and $h_p$ is a bandwidth parameter.
Furthermore, the asymptotic covariance matrix $\Omega_{\nong}$ of the unprewhitened estimator $\hat{\beta}(I_{p})$ can be estimated by
\begin{equation}
\hat{\Omega}_{\nong}=\hat S^{-1}_{xx}(I_{p})\{I_{m}+\Lambda\hat\beta(I_{p})\hat\beta^{T}(I_{p})\}^{-1}\hat{V}_{g}\{I_{m}+\hat\beta(I_{p})\hat\beta^{T}(I_{p})\Lambda\}^{-1}\hat S^{-1}_{xx}(I_{p}).
\end{equation}

\begin{condition}
\label{asp:kernel}
(i) The kernel function $\mathcal{K}:[0, \infty) \to [-1, 1]$ satisfies $\mathcal{K}(0) = 1$, $\mathcal{K}(x) \to 1$ as $x \to 0$, and $\mathcal{K}(x) = 0$ for $x > 1$, and the bandwidth $h_p$ satisfies $h_p \to \infty$ and $h_p=o(p^{1/2d})$ as $p \to \infty$; and
(ii) the collections $\{\|X_i\|^{8+4\delta}: i \ge 1\}$ and $\{\|\mathcal{E}_i\|^{8+4\delta}: i \ge 1\}$ are uniformly integrable for a constant $0<\delta<1$.
\end{condition}

\begin{theorem}
\label{thm:HAC-estimator}
Under Conditions \ref{asp:mixing}--\ref{asp:trace}, \ref{asp:moment-nonGaussian-general}, and \ref{asp:kernel} for Model \eqref{eq:MER-y}--\eqref{eq:MER-e}, we have $\hat{V}_{g}\to_{\pr}V_{g}$ and thus $\hat\Omega_{\nong}\to_{\pr}\Omega_{\nong}$.
\end{theorem}

Note that the consistency of the proposed kernel estimator does not require the exact Gaussian or non-Gaussian moment configurations in Theorem \ref{thm:TLS-estimator} or \ref{thm:TLS-estimator-nonGaussian}, respectively.
This makes $\hat\Omega_{\nong}$ robust to unknown distributions in practice.

\section{Simulation studies}
\label{sec:simu}

To further demonstrate the difference between $\hat{\beta}(I_{p})$ and $\hat{\beta}(\hat{\Sigma}^{-1})$ in terms of statistical efficiencies and computational costs, we conducted a simulation study as follows.
We fixed the observation sample size at $p=572$ and the ensemble sample size at $n=56,223,892,\infty$ to meet the practical situation in \cite{LCY21}, where the case of $(p,n)=(572,223)$ serves the baseline setting.
The cases with $n=56,223,892$ reflected insufficient, moderate, and rich but unrealistic computational resources.
The ideal case of $n=\infty$ reflected the situation of unlimited computational resources so that $\Sigma$ is essentially known.
The dimension of the covariates was fixed to $m=1$ and the regression coefficient was set to $\beta_{1}=2$ for ease of explanation. 

We adopted the setting in \cite{CZZ10-MR2676885} to construct the error covariance matrix as $\Sigma=(\sigma_{ij})$, where $\sigma_{ij}=\sigma_{i}^{2}$ for $1\leq i=j\leq p$ and $\sigma_{ij}=\rho|i-j|^{-(\alpha+1)}\sigma_{i}\sigma_{j}$ for $1\leq i\neq j\leq p$.
Here, $\sigma_{1}^{2},\dots,\sigma_{p}^{2}$ were independently generated from the uniform distribution over $(0.02,0.18)$ and were kept fixed throughout of the simulation.
We took the decaying parameter $\alpha=0.1,0.3,0.5$ and the scaling parameter $\rho=0.2,0.4,0.6$ to represent various dependence structures.
Following Example \ref{cex:beat}, the unobservable true covariate vector was fixed as $x_{1}=(x_{11},\dots,x_{1p})^{T}$, where $x_{1i}=\sigma_{i}^{2}/\{1-(1+\beta_{1}^{2})\sigma_{i}^{2}\}^{1/2}$.
The random errors $\varepsilon,u_{1},v_{1},\dots,v_{n}$ in \eqref{eq:MER-y}--\eqref{eq:MER-e} were independently generated from $N(0,\Sigma)$ so that $\Lambda=1$, where $v_{1},\dots,v_{n}$ represents the replicate errors used to estimate $\Sigma$.
For the case of $n=\infty$, we simply took $\hat{\Sigma}=\Sigma$.
For the cases with $n<\infty$, we employed the tapering estimator described in \cite{CZZ10-MR2676885}, namely $\hat{\Sigma}=(w_{ij}\tilde{\sigma}_{ij})$ for weights $w_{ij}=(2/k)\{(k-|i-j|)_{+}-(k/2-|i-j|)_{+}\}$, bandwidth $k=\lfloor n^{1/(2\alpha+1)}\rfloor$, ensemble covariance matrix $\tilde{\Sigma}=(\tilde{\sigma}_{ij})=n^{-1}\sum_{i=1}^{n}(v_{i}-\bar{v})(v_{i}-\bar{v})^{T}$, and ensemble mean $\bar{v}=n^{-1}\sum_{i=1}^{n}v_{i}$.

\begin{table}[ht]
\centering
\caption{Empirical coverage rates and lengths (in parentheses) of the confidence intervals based on $\hat{\beta}(I_{p})$ and $\hat{\beta}(\hat{\Sigma}^{-1})$ over 1000 simulation replications with covariate dimension $m=1$, regression coefficient $\beta_{1}=2$, observation sample size $p=572$, and nominal coverage rate $95\%$.}
\label{tab:CR_length_combined}
\resizebox{\textwidth}{!}{
\begin{tabularx}{\linewidth}{c r *{6}{Y}}
\toprule
 & & 
\multicolumn{2}{c}{$\alpha=0.1$} & 
\multicolumn{2}{c}{$\alpha=0.3$} & 
\multicolumn{2}{c}{$\alpha=0.5$} \\
\cmidrule(lr){3-4} \cmidrule(lr){5-6} \cmidrule(lr){7-8}
$\rho$ & $n$ & $\hat{\beta}(I_{p})$ & $\hat{\beta}(\hat{\Sigma}^{-1})$ & $\hat{\beta}(I_{p})$ & $\hat{\beta}(\hat{\Sigma}^{-1})$ & $\hat{\beta}(I_{p})$ & $\hat{\beta}(\hat{\Sigma}^{-1})$ \\
\midrule
\multirow{8}{*}{0.2} 
& \multirow{2}{*}{56} & 94.6\% & 8.8\% & 94.5\% & 75.2\% & 94.7\% & 85.4\% \\
& & (1.002) & (63.612) & (0.922) & (0.859) & (0.886) & (0.916) \\
\cmidrule(lr){2-8}
& \multirow{2}{*}{223} & 93.7\% & 14.9\% & 93.2\% & 86.8\% & 93.3\% & 90.2\% \\
& & (1.018) & (30.258) & (0.928) & (1.012) & (0.889) & (1.014) \\
\cmidrule(lr){2-8}
& \multirow{2}{*}{892} & 95.5\% & 75.1\% & 95.4\% & 92.5\% & 95.1\% & 94.9\% \\
& & (1.048) & (0.991) & (0.948) & (1.098) & (0.905) & (1.073) \\
\cmidrule(lr){2-8}
& \multirow{2}{*}{$\infty$} & 95.6\% & 94.7\% & 95.6\% & 95.2\% & 95.8\% & 95.1\% \\
& & (1.035) & (1.359) & (0.958) & (1.226) & (0.915) & (1.144) \\
\cmidrule(lr){1-8}
\multirow{8}{*}{0.4} 
& \multirow{2}{*}{56} & 93.0\% & 8.0\% & 93.6\% & 75.5\% & 94.1\% & 85.5\% \\
& & (1.182) & (217.680) & (1.058) & (0.934) & (0.998) & (0.984) \\
\cmidrule(lr){2-8}
& \multirow{2}{*}{223} & 92.7\% & 16.1\% & 92.6\% & 87.3\% & 92.8\% & 90.6\% \\
& & (1.221) & (1510.977) & (1.078) & (1.100) & (1.011) & (1.096) \\
\cmidrule(lr){2-8}
& \multirow{2}{*}{892} & 95.0\% & 75.4\% & 94.5\% & 92.1\% & 94.5\% & 94.7\% \\
& & (1.277) & (1.071) & (1.114) & (1.190) & (1.039) & (1.158) \\
\cmidrule(lr){2-8}
& \multirow{2}{*}{$\infty$} & 95.2\% & 94.8\% & 95.2\% & 95.0\% & 95.2\% & 95.1\% \\
& & (1.270) & (1.454) & (1.135) & (1.326) & (1.059) & (1.235) \\
\cmidrule(lr){1-8}
\multirow{8}{*}{0.6} 
& \multirow{2}{*}{56} & 92.4\% & 8.6\% & 92.5\% & 72.1\% & 93.4\% & 83.9\% \\
& & (1.359) & (23.415) & (1.195) & (0.707) & (1.114) & (0.697) \\
\cmidrule(lr){2-8}
& \multirow{2}{*}{223} & 92.0\% & 17.1\% & 92.1\% & 86.0\% & 92.3\% & 91.3\% \\
& & (1.415) & (3782.445) & (1.225) & (0.835) & (1.134) & (0.795) \\
\cmidrule(lr){2-8}
& \multirow{2}{*}{892} & 94.5\% & 73.0\% & 94.3\% & 91.8\% & 94.3\% & 94.4\% \\
& & (1.497) & (0.802) & (1.276) & (0.895) & (1.173) & (0.833) \\
\cmidrule(lr){2-8}
& \multirow{2}{*}{$\infty$} & 95.0\% & 95.2\% & 95.1\% & 95.5\% & 95.0\% & 95.1\% \\
& & (1.493) & (1.089) & (1.306) & (0.977) & (1.200) & (0.871) \\
\bottomrule
\end{tabularx}}
\end{table}

Table \ref{tab:CR_length_combined} reports 
the empirical coverage rates and lengths of the nominal 95\% confidence intervals constructed using the unprewhitened estimator $\hat{\beta}(I_{p})$ and the prewhitened estimator $\hat{\beta}(\hat{\Sigma}^{-1})$, respectively, base on $10000$ simulation replications.
The confidence intervals based on $\hat{\beta}(I_{p})$ were constructed by using the plug-in estimator $\hat{\Omega}$ of $\Omega$ as described below Theorem \ref{thm:TLS-estimator}.
The confidence intervals based on $\hat{\beta}(\hat{\Sigma}^{-1})$ were constructed similarly as in Theorem \ref{thm:WTLS-estimator}.
The lengths for the two confidence intervals were defined as $2z_{0.025}(\hat{\Omega}/p)^{1/2}$ and $2z_{0.025}(\hat{\Omega}^{**}/p)^{1/2}$, respectively, where $z_{0.025}\approx1.96$ is the $2.5\%$ upper quantile of $N(0,1)$.

The unprewhitened confidence intervals had an empirical coverage rate close to the nominal level $95\%$ 
even when the ensemble sample size $n$ was limited at $56$ for all parameter combinations of $(\alpha,\rho)$ for the error covariance matrix $\Sigma$.
The empirical coverage rates became closer to $95\%$ as $n$ increased.
This shows that the required ensemble sample size of the unprewhitened confidence interval can be much smaller than the benchmark size $n=223$ in \cite{LCY21}, which is computationally efficient and is expected by the analysis in Section \ref{sec:comp-cost} with a logarithmic rate on the ensemble size $n$ such that $\log p=o(n)$.

In contrast, the prewhitened confidence intervals had severe under-coverage issue.
In particular, when the underlying dependence was long range with $\alpha=0.1$, the coverage rates of the prewhitened confidence intervals were all below $80\%$ except for the ideal case of $n=\infty$.
In other words, the prewhitened method produced less satisfactory confidence intervals than the unprewhitened method, unless a much larger ensemble size is required as expected by the analysis in Section \ref{sec:comp-cost} with a much stronger polynomial rate condition $p^{\min\{2,(2\alpha+1)/2\alpha\}}=o(n)$.
The under-coverage issue was reduced as the underlying dependence became weaker with a larger decaying parameter $\alpha$ and a smaller scaling parameter $\rho$.

When $n=\infty$, the unprewhitened confidence intervals were narrower than the prewhitened counterparts for $\rho=0.2$ and $0.4$.
{In this case, the prewhitened method was} not superior to the unprewhitened method even if the computational resources were unlimited, as expected by the analysis in Example \ref{cex:beat}.
The prewhitened confidence intervals were narrower than the unprewhitened counterparts when $\rho=0.6$ and $n$ is larger than the benchmark size ($n=223$).
However, {for $n$ smaller than or equal to the benchmark size,} the prewhitened method was not as stable as the unprewhitened method and might produce an extremely wide confidence interval, which was unsuitable for real world applications.

In summary, each of the two methods might be more statistically efficient than the other with a narrower confidence interval at certain settings. There is no guarantee that the prewhitened estimator would lead to a more efficient estimation.
However, it always uses much more computational expenses and {might} not be able to produce  reliable {confidence intervals} due to the limited ensemble sizes encountered in practice.
These confirmed  the theoretical analysis in Sections \ref{sec:stat-efficiency} and \ref{sec:comp-cost}.
In contrast, the behavior of the unprewhitened confidence intervals was much more stable with a much smaller ensemble sample size required to ensure a valid statistical inference.
Therefore, it is safe to conclude that the unprewhitened method is a more practically desirable choice than the prewhitened method.
%
%

%
%

\bibliographystyle{plainnat} 
\bibliography{EIV} 

\appendix
\section{Useful lemmas of matrix normal distributions}
To facilitate the derivation of asymptotic results, we invoke some useful lemmas of matrix normal distributions from \cite{KvR05-MR2162145}.
We first introduce additional notations of moments of random matrices.
Let $A^{\otimes k}=A\otimes\cdots\otimes A$ be the $k$-th Kroneckerian power of $A$.
For an $n\times m$ random matrix $X$, let $\mathfrak{M}_{k}(X)=E\{\vecc X(\vecc ^{T}X)^{\otimes k-1}\}\in\mathbb{R}^{nm\times (nm)^{k-1}}$ be its $k$-th moment and $\var(X)=\mathfrak{M}_{2}\{X-E(X)\}$ be its covariance matrix.
For a general matrix $M\in\mathbb{R}^{n\times m}$ and two positive semidefinite matrices $\Sigma\in\mathbb{R}^{n\times n}$ and $\Psi\in\mathbb{R}^{m\times m}$, a random matrix $X$ has matrix normal distribution denoted by $X\sim N(M,\Sigma,\Psi)$ if and only if $\vecc X\sim N(\vecc M,\Psi\otimes\Sigma)$, where $\vecc X$ is the vectorization of $X$.
Let $K_{n,m}\in\mathbb{R}^{nm\times nm}$ be the commutation matrix such that $K_{n,m}\vecc A=\vecc (A^{T})$ for every $n\times m$ matrix $A$.
Let $N_{n}=(I_{n^{2}}+K_{n,n})/2$.

\begin{lemma}
\label{lem:KvR05-Kronecker}
(i) Let $A:p\times q$, $B:q\times r$, and $C:r\times s$.
Then,
\begin{equation}
\vecc(ABC)=(C^{T}\otimes A)\vecc B.
\end{equation}
(ii) Let $A:p\times q$, $B:q\times w$, $C:r\times s$, and $D:s\times t$.
Then,
\begin{equation}
(A\otimes C)(B\otimes D)=(AB)\otimes(CD).
\end{equation}
(iii) Let $A:p\times q$ and $B:r\times s$.
Then,
\begin{equation}
A\otimes B=K_{p,r}(B\otimes A)K_{s,q}.
\end{equation}
(iv) We have $K_{p,q}K_{q,p}=I_{pq}$.
\end{lemma}
\begin{proof}
(i), (ii), (iii), and (iv) are, respectively, Propositions 1.3.14(ii), 1.3.12(vi), 1.3.12(viii), and 1.3.10(ii) in \cite{KvR05-MR2162145}.
\end{proof}

\begin{lemma}
\label{lem:Thm229-KvR05}
Let $X\sim N(O_{n\times m},\Sigma,\Psi)$ and $A$ be nonrandom and symmetric.
Then,
\begin{equation}
\mathbb{E}(XAX^{T})=\tr(\Psi A)\Sigma\quad\text{and}\quad\var(XAX^{T})=2\tr(\Psi A\Psi A)N_{n}(\Sigma\otimes\Sigma).
\end{equation}
\end{lemma}
\begin{proof}
See Theorem 2.2.9 in \cite{KvR05-MR2162145}.
\end{proof}

\section{Proof of Theorem \ref{thm:TLS-estimator}, Corollary \ref{cor:TLS-estimator-general}, and Theorem \ref{thm:TLS-estimator-nonGaussian-general}}
\label{pf:TLS}
We only prove Theorem \ref{thm:TLS-estimator}, as Corollary \ref{cor:TLS-estimator-general} is a direct consequence of Theorem \ref{thm:TLS-estimator}, and Theorem \ref{thm:TLS-estimator-nonGaussian-general} can be immediately implied by the asymptotic expansion \eqref{eq:asymptotic-expression} below.

(i) Observe that by Condition \ref{asp:moment}(ii), the first four moments of $\mathcal{E}$, conditioning on $X$, are the same as those of $N(O_{p\times(m+1)},\Sigma,\Psi)$, or the first four moments of $\mathcal{E}^{T}$, conditioning on $X$, are the same as those of $N(O_{(m+1)\times p},\Psi,\Sigma)$.
Let $\mathcal{B}=(I_{m},\beta)$ so that Model \eqref{eq:MER-y}--\eqref{eq:MER-e} can be rewritten as $W=X\mathcal{B}+\mathcal{E}$.
For simplicity, we shall write $S_{ww}=S_{ww}(I_{p})$ in this proof when doing so does not cause ambiguity.
Let $S_{xe}=X^{T}\mathcal{E}/p$, $S_{ex}=\mathcal{E}^{T}X/p$, and $S_{ee}=\mathcal{E}^{T}\mathcal{E}/p$.
Then,
\begin{equation}
\label{eq:G-decomposition}
S_{ww}=\mathcal{B}^{T}S_{xx}\mathcal{B}+\mathcal{B}^{T}S_{xe}+S_{ex}\mathcal{B}+S_{ee}.
\end{equation}

We first prove the consistency part of (i).
By Conditions \ref{asp:moment}(i) and \ref{asp:mixing}, it follows from the $L_{1}$ law of large numbers for random fields \citep[Theorem 3]{JP09-MR2525996} that $S_{ww}-\mathbb{E}S_{ww}\to0$ in $L_{1}$ and thus in probability.
By Condition \ref{asp:trace}(i), we have $ES_{xx}\to Q_{0}$.
By Conditions \ref{asp:moment}(ii) and \ref{asp:trace}(ii), and Lemma \ref{lem:Thm229-KvR05}, we have $ES_{xe}=0$ and $ES_{ee}=p^{-1}\tr(\Sigma)\Psi\to\tau_{1}\Psi$.
Consequently,
\begin{equation}
S_{ww}\overset{\pr}{\to}\mathcal{B}^{T}Q_{0}\mathcal{B}+\tau_{1}\Psi,
\end{equation}
which implies that
\begin{equation}
\label{eq:many-Q-consistency}
S_{zz}\overset{\pr}{\to}Q_{0}+\tau_{1}\Lambda,\quad\frac{1}{\lambda_{\max}(S^{-1}_{ww}\Psi)}\overset{\pr}{\to}\tau_{1},\quad\text{and}\quad S_{zy}\overset{\pr}{\to}Q_{0}\beta,
\end{equation}
where we have adopted the fact that
\begin{equation}
\lambda_{\max}((\mathcal{B}^{T}Q_{0}\mathcal{B}+\tau_{1}\Psi)^{-1}\Psi)=\max_{x\neq0}\frac{x^{T}\Psi x}{x^{T}(\mathcal{B}^{T}Q_{0}\mathcal{B}+\tau_{1}\Psi)x}=\frac{1}{\tau_{1}},
\end{equation}
and the maximum is attained at $x=(-\beta^{T},1)^{T}=:\mathcal{B}^{\perp}$ with $\mathcal{B}\mathcal{B}^{\perp}=O_{m\times1}$.
The both consistency results in \eqref{eq:TLS-estimator-asymptotic-normality} and $\hat{S}_{xx}\to_{\pr}Q_{0}$ are direct consequences of \eqref{eq:many-Q-consistency}.
Note also that $S_{ww}$ is nonsingular with probability approaching to one, as by Condition \ref{asp:trace},
\begin{equation}
\det(S_{ww})\overset{\pr}{\to}[\beta^{T}Q_{0}\{Q^{-1}_{0}-(Q_{0}+\tau_{1}\Lambda)^{-1}\}Q_{0}\beta+\tau_{1}]\det(Q_{0}+\tau_{1}\Lambda)>0,
\end{equation}
which completes the proof of the consistency part.

We now prove the asymptotic normality part of (i).
Observe that with probability approaching to one, $\hat{\beta}=\hat{\beta}(I_{p})$ satisfies
\begin{equation}
(S_{ww}-\hat{\lambda}\Psi)\begin{pmatrix}-\hat{\beta}\\
1
\end{pmatrix}=0,
\label{eq:estimating-equation}
\end{equation}
where $\hat{\lambda}=1/\lambda_{\max}(S^{-1}_{ww}\Psi)$.
Indeed,
\begin{equation}
(S_{ww}-\hat{\lambda}\Psi)\begin{pmatrix}-\hat{\beta}\\
1
\end{pmatrix}=\begin{pmatrix}S_{zz}-\hat{\lambda}\Lambda & S_{zy}\\
S_{yz} & S_{yy}-\hat{\lambda}
\end{pmatrix}\begin{pmatrix}-\hat{\beta}\\
1
\end{pmatrix}=\begin{pmatrix}-(S_{zz}-\hat{\lambda}\Lambda)\hat{\beta}+S_{zy}\\
-S_{yz}\hat{\beta}+S_{yy}-\hat{\lambda}
\end{pmatrix},
\label{eq:estimating-equation-1}
\end{equation}
where the top block is zero due to the definition of $\hat{\beta}$.
For the bottom block, we notice that
\begin{equation}
\det(S_{ww}-\hat{\lambda}\Psi)=\det(S_{ww})\det(I_{m+1}-\hat{\lambda}S^{-1}_{ww}\Psi)=0,
\label{eq:estimating-equation-2}
\end{equation}
while on the other hand,
\begin{align}
\det(S_{ww}-\hat{\lambda}\Psi)&=\{S_{yy}-\hat{\lambda}-S_{yz}(S_{zz}-\hat{\lambda}\Lambda)^{-1}S_{zy}\}\det(S_{zz}-\hat{\lambda}\Lambda)\notag\\&=(-S_{yz}\hat{\beta}+S_{yy}-\hat{\lambda})\det(\hat{S}_{xx}).
\label{eq:estimating-equation-3}
\end{align}
As $\hat{S}_{xx}\to_{\pr}Q_{0}$, we have $\hat{S}_{xx}$ is nonsingular with probability approaching to one.
By \eqref{eq:estimating-equation-2} and \eqref{eq:estimating-equation-3}, this implies $-S_{yz}\hat{\beta}+S_{yy}-\hat{\lambda}=0$ and thus  \eqref{eq:estimating-equation} holds with probability approaching to one.

Let $\mathcal{L}=(I_{m},\Lambda\beta)$ and $\mathcal{H}=(S_{ww}-\hat{\lambda}\Psi)(I_{m},O_{m\times1})^{T}$ so that $\mathcal{L}\Psi\mathcal{B}^{\perp}=O_{m\times1}$.
Let $S_{we}=W^{T}\mathcal{E}/p$.
Then, it follows from \eqref{eq:estimating-equation} that with probability approaching to one,
\begin{equation}
(S_{ww}-\hat{\lambda}\Psi)\mathcal{B}^{\perp}=(S_{ww}-\hat{\lambda}\Psi)\begin{pmatrix}\hat{\beta}-\beta\\
0
\end{pmatrix}=\mathcal{H}(\hat{\beta}-\beta),
\end{equation}
and thus
\begin{equation}
\label{eq:Sww-zero-mean}
\mathcal{L}\mathcal{H}(\hat{\beta}-\beta)=\mathcal{L}S_{ww}\mathcal{B}^{\perp}=\mathcal{L}(S_{we}-ES_{we})\mathcal{B}^{\perp},
\end{equation}
where
\begin{equation}
\mathcal{L}\mathcal{H}\overset{\pr}{\to}(I_{m},\Lambda\beta)\mathcal{B}^{T}Q_{0}\mathcal{B}\begin{pmatrix}I_{m}\\
0
\end{pmatrix}=(I_{m}+\Lambda\beta\beta^{T})Q_{0},
\end{equation}
which is nonsingular as $\det(I_{m}+\Lambda\beta\beta^{T})=1+\beta^{T}\Lambda\beta>0$.
Consequently,
\begin{equation}
\hat{\beta}-\beta=Q^{-1}_{0}(I_{m}+\Lambda\beta\beta^{T})^{-1}\mathcal{L}(S_{we}-ES_{we})\mathcal{B}^{\perp}\{1+o_{\pr}(1)\}.
\label{eq:asymptotic-expression}
\end{equation}

By Condition \ref{asp:moment}(ii), we have
\begin{equation}
\var(S_{we})=\var(S_{ee})+\var(\mathcal{B}^{T}S_{xe}).
\label{eq:G-decomposition-var}
\end{equation}
By Lemma \ref{lem:Thm229-KvR05} and Condition \ref{asp:trace}(ii),
\begin{equation}
\var(p^{1/2}S_{ee})=E\var(p^{1/2}S_{ee}\mid X)=\frac{2}{p}\tr(\Sigma^{2})N_{m+1}(\Psi\otimes\Psi)\to2\tau_{2}N_{m+1}(\Psi\otimes\Psi).
\label{eq:G-decomposition-var-1}
\end{equation}
By Condition \ref{asp:moment}(ii), the first two moments of $\mathcal{B}^{T}X^{T}\mathcal{E}$, conditioning on $X$, are the same as those of $N(0,\mathcal{B}^{T}X^{T}\Sigma X\mathcal{B},\Psi)$.
Then, by Condition \ref{asp:trace}(i),
\begin{equation}
\var(p^{1/2}\mathcal{B}^{T}S_{xe})=E\var(p^{1/2}\mathcal{B}^{T}S_{xe}\mid X)=\frac{1}{p}\Psi\otimes(\mathcal{B}^{T}X^{T}\Sigma X\mathcal{B})\to\Psi\otimes(\mathcal{B}^{T}Q_{1}\mathcal{B}).
\label{eq:G-decomposition-var-2}
\end{equation}
By combining \eqref{eq:G-decomposition-var}--\eqref{eq:G-decomposition-var-2}, we have
\begin{equation}
\label{eq:Qzz-var}
\var(p^{1/2}S_{we})\to2\tau_{2}N_{m+1}(\Psi\otimes\Psi)+\Psi\otimes(\mathcal{B}^{T}Q_{1}\mathcal{B}).
\end{equation}
Note also that by Lemma \ref{lem:KvR05-Kronecker},
\begin{align}
&[(\mathcal{B}^{\perp})^{T}\otimes\{Q^{-1}_{0}(\mathcal{L}\mathcal{B}^{T})^{-1}\mathcal{L}\}]K_{m+1,m+1}(\Psi\otimes\Psi)[\mathcal{B}^{\perp}\otimes\{\mathcal{L}^{T}(\mathcal{B}\mathcal{L}^{T})^{-1}Q^{-1}_{0}\}]\notag\\&=K_{1,m}[\{Q^{-1}_{0}(\mathcal{L}\mathcal{B}^{T})^{-1}\mathcal{L}\}\otimes(\mathcal{B}^{\perp})^{T}](\Psi\otimes\Psi)[\mathcal{B}^{\perp}\otimes\{\mathcal{L}^{T}(\mathcal{B}\mathcal{L}^{T})^{-1}Q^{-1}_{0}\}]\notag\\&=K_{1,m}\{Q^{-1}_{0}(\mathcal{L}\mathcal{B}^{T})^{-1}\mathcal{L}\Psi\mathcal{B}^{\perp}\}\otimes\{(\mathcal{B}^{\perp})^{T}\Psi\mathcal{L}^{T}(\mathcal{B}\mathcal{L}^{T})^{-1}Q^{-1}_{0}\}=O_{m\times m}.
\label{eq:beta-var-pre}
\end{align}
Then, by combining \eqref{eq:asymptotic-expression} and \eqref{eq:Qzz-var}--\eqref{eq:beta-var-pre}, we can conclude that
\begin{align}
&\var\{p^{1/2}(\hat{\beta}-\beta)\}\notag\\&=[(\mathcal{B}^{\perp})^{T}\otimes\{Q^{-1}_{0}(\mathcal{L}\mathcal{B}^{T})^{-1}\mathcal{L}\}]\var(p^{1/2}S_{we})[\mathcal{B}^{\perp}\otimes\{\mathcal{L}^{T}(\mathcal{B}\mathcal{L}^{T})^{-1}Q^{-1}_{0}\}]\{1+o(1)\}\notag\\&\to[(\mathcal{B}^{\perp})^{T}\otimes\{Q^{-1}_{0}(\mathcal{L}\mathcal{B}^{T})^{-1}\mathcal{L}\}]\{\Psi\otimes(\tau_{2}\Psi+\mathcal{B}^{T}Q_{1}\mathcal{B})\}[\mathcal{B}^{\perp}\otimes\{\mathcal{L}^{T}(\mathcal{B}\mathcal{L}^{T})^{-1}Q^{-1}_{0}\}]\notag\\&=(1+\beta^{T}\Lambda\beta)Q^{-1}_{0}\{\tau_{2}(\mathcal{L}\mathcal{B}^{T})^{-1}\mathcal{L}\Psi\mathcal{L}^{T}(\mathcal{B}\mathcal{L}^{T})^{-1}+Q_{1}\}Q^{-1}_{0}=\Omega.
\label{eq:beta-var}
\end{align}
By Conditions \ref{asp:moment}(i) and \ref{asp:mixing}, and the central limit theorem for random fields \citep[Corollary 1]{JP09-MR2525996}, the entries of $p^{1/2}(S_{we}-ES_{we})$ on or below the diagonal admit the asymptotic normality with zero means and the asymptotic covariance matrix given in \eqref{eq:Qzz-var}.
Therefore, by the asymptotic expansion in \eqref{eq:asymptotic-expression} and the asymptotic covariance matrix in \eqref{eq:beta-var}, the asymptotic normality result in \eqref{eq:TLS-estimator-asymptotic-normality} follows.
This completes the proof of (i).

(ii) For $\hat{\tau}_{2}\to_{\pr}\tau_{2}$, it suffices to notice that $p^{-1/2}\tr^{1/2}\{(\hat{\Sigma}-\Sigma)^{2}\}\leq\|\hat{\Sigma}-\Sigma\|=o_{\pr}(1)$ and then apply triangle inequality for the Frobenius norm.
For $\hat{S}_{xx}(\hat{\Sigma})\to_{\pr}Q_{1}$, it can be shown that $S_{zz}(\Sigma)-\hat{\lambda}(\Sigma)\Lambda\to_{\pr}Q_{1}$ and that $\|S_{zz}(\hat{\Sigma})-S_{zz}(\Sigma)\|\leq\|S_{zz}(I_{p})\|\|\hat{\Sigma}-\Sigma\|=o_{\pr}(1)$.
Then, we can apply Lemmas \ref{lem:PerturbationOfMatrixInversion} and \ref{lem:WeylPerturbationTheorem} below to find that $|\lambda_{\max}(S_{ww}^{-1}(\hat{\Sigma})\Psi)-\lambda_{\max}(S_{ww}^{-1}(\Sigma)\Psi)|\leq\|\Psi\|\|S^{-1}_{ww}(\hat{\Sigma})-S^{-1}_{ww}(\Sigma)\|=o_{\pr}(1)$.
This completes the proof of (ii).

\section{Proof of Theorem \ref{thm:PerturbationOfTheTLSEstimator}}
We first introduce two lemmas concerning the perturbation bounds for the matrix inversion and the eigenvalues, which can be found in \citet[Theorem 8.1.2]{WWQ18-MR3793648} and \citet[Theorem VI.2.1]{Bha97-MR1477662} respectively.

\begin{lemma}
\label{lem:PerturbationOfMatrixInversion}
Let $A$, $B$ be $n\times n$ real matrices.
Assume that $A$ is invertible and $\Delta:=\|A^{-1}\|\|B-A\|<1$.
Then $B$ is also invertible and satisfies
\begin{equation}
\label{eq:lem-PerturbationOfMatrixInversion}
\|B^{-1}\|\leq\frac{1}{1-\Delta}\|A^{-1}\|\quad\text{and}\quad
\|B^{-1}-A^{-1}\|\leq\frac{\Delta}{1-\Delta}\|A^{-1}\|.
\end{equation}
\end{lemma}

\begin{lemma}
\label{lem:WeylPerturbationTheorem}
Let $A$, $B$ be $n\times n$ real symmetric matrices with eigenvalues $\lambda_{1}^{\downarrow}(A)\geq\cdots\geq\lambda_{n}^{\downarrow}(A)$ and $\lambda_{1}^{\downarrow}(B)\geq\cdots\geq\lambda_{n}^{\downarrow}(B)$ respectively.
Then
\begin{equation}
\max_{1\leq j\leq n}\vert\lambda_{j}^{\downarrow}(B)-\lambda_{j}^{\downarrow}(A)\vert\leq\|B-A\|.
\end{equation}
\end{lemma}

Let $\Delta=\Delta(B,A)$ and $\hat{\Delta}=\ub(\Delta(\hat{S}_{xx}(B),\hat{S}_{xx}(A)))$.
Returning to the proof of Theorem \ref{thm:PerturbationOfTheTLSEstimator}, as $\Delta\leq\hat{\Delta}<1$, it follows from Lemma \ref{lem:WeylPerturbationTheorem} that
\begin{equation}
\Big\vert\frac{\lambda_{\min}(B)}{\lambda_{\min}(A)}-1\Big\vert\leq\frac{\|B-A\|}{\lambda_{\min}(A)}=\Delta<1.
\end{equation}
This implies that $\lambda_{\min}(B)\in(0,2\lambda_{\min}(A))$ and therefore $B$ is positive definite.

We first analyze the perturbation bounds for the components of $\hat{S}_{xx}(A)$.
Observe that $B = A^{1/2}\{I_{p} + A^{-1/2}(B-A)A^{-1/2}\}A^{1/2}$.
Since $\|A^{-1/2}(B-A)A^{-1/2}\| \leq \|A^{-1}\| \|B-A\| = \Delta$,
\begin{equation}
(1-\Delta) A \leq B \leq (1+\Delta) A,
\end{equation}
where $A\leq B$ denotes $B-A$ is positive semidefinite.
Pre- and post-multiplying by $W^T/p^{1/2}$ and $W/p^{1/2}$, respectively, yield
\begin{equation}
\label{eq:thm-PerturbationOfTheTLSEstimator-Sww}
(1-\Delta) S_{ww}(A) \leq S_{ww}(B) \leq (1+\Delta) S_{ww}(A),
\end{equation}
which implies that $S_{ww}(B)$ is positive definite.
Note also that
\begin{equation}
\label{eq:thm-PerturbationOfTheTLSEstimator-Rayleigh}
\hat{\lambda}(A)=\min_{x:x^{T}\Psi x\neq0}\frac{x^{T}S_{ww}(A)x}{x^{T}\Psi x}.
\end{equation}
Consequently, by \eqref{eq:thm-PerturbationOfTheTLSEstimator-Sww} and \eqref{eq:thm-PerturbationOfTheTLSEstimator-Rayleigh}, we have
\begin{equation}
(1-\Delta)\hat{\lambda}(A) \leq \hat{\lambda}(B) \leq (1+\Delta)\hat{\lambda}(A),
\end{equation}
or equivalently,
\begin{equation}
\label{eq:thm-PerturbationOfTheTLSEstimator-lambda-diff}
|\hat{\lambda}(B) - \hat{\lambda}(A)| \leq \Delta \hat{\lambda}(A).
\end{equation}
Similarly, we have
\begin{equation}
\label{eq:thm-PerturbationOfTheTLSEstimator-block-diff}
\|S_{zz}(B) - S_{zz}(A)\| \leq \Delta \|S_{zz}(A)\| \quad \text{and} \quad \|S_{zy}(B) - S_{zy}(A)\| \leq \Delta \|S_{zz}(A)\|^{1/2} S_{yy}^{1/2}(A).
\end{equation}

Next, we derive the perturbation bound for $\hat{S}_{xx}(A)$.
By combining \eqref{eq:thm-PerturbationOfTheTLSEstimator-lambda-diff} and \eqref{eq:thm-PerturbationOfTheTLSEstimator-block-diff}, we have
\begin{align}
\label{eq:thm-PerturbationOfTheTLSEstimator-Sxx-diff}
\|\hat{S}_{xx}(B) - \hat{S}_{xx}(A)\| &\leq \|S_{zz}(B) - S_{zz}(A)\| + |\hat{\lambda}(B) - \hat{\lambda}(A)| \|\Lambda\| \notag\\
&\leq \Delta \|S_{zz}(A)\| + \Delta \hat{\lambda}(A) \|\Lambda\|.
\end{align}
Multiplying \eqref{eq:thm-PerturbationOfTheTLSEstimator-Sxx-diff} by $\|\hat{S}_{xx}^{-1}(A)\| = 1/\lambda_{\min}(\hat{S}_{xx}(A))$, we have
\begin{equation}
\|\hat{S}_{xx}^{-1}(A)\| \|\hat{S}_{xx}(B) - \hat{S}_{xx}(A)\| \leq \frac{\|S_{zz}(A)\| + \hat{\lambda}(A) \|\Lambda\|}{\lambda_{\min}(\hat{S}_{xx}(A))} \Delta = \hat{\Delta} < 1.
\end{equation}
By Lemma \ref{lem:PerturbationOfMatrixInversion}, this implies that $\hat{S}_{xx}(B)$ is positive definite, and that
\begin{equation}
\label{eq:thm-PerturbationOfTheTLSEstimator-gamma-B-inverse}
\|\hat{S}_{xx}^{-1}(B)\| \leq \frac{1}{1-\hat{\Delta}} \|\hat{S}_{xx}^{-1}(A)\|\quad\text{and}\quad\|\hat{S}_{xx}^{-1}(B) - \hat{S}_{xx}^{-1}(A)\| \leq \frac{\hat{\Delta}}{1-\hat{\Delta}} \|\hat{S}_{xx}^{-1}(A)\|.
\end{equation}

We now derive the perturbation bound for $\hat{\beta}(A)$.
By combining \eqref{eq:thm-PerturbationOfTheTLSEstimator-block-diff} and \eqref{eq:thm-PerturbationOfTheTLSEstimator-gamma-B-inverse}, we have
\begin{align}
&\label{eq:thm-PerturbationOfTheTLSEstimator-beta-diff}\|\hat{\beta}(B)-\hat{\beta}(A)\|=\|\hat{S}^{-1}_{xx}(B)S_{zy}(B)-\hat{S}^{-1}_{xx}(A)S_{zy}(A)\|\notag\\&\leq\|\hat{S}^{-1}_{xx}(B)\|\|S_{zy}(B)-S_{zy}(A)\|+\|\hat{S}^{-1}_{xx}(B)-\hat{S}^{-1}_{xx}(A)\|\|S_{zy}(A)\|\notag\\&\leq\frac{1}{1-\hat{\Delta}}\|\hat{S}^{-1}_{xx}(A)\|\Delta\|S_{zz}(A)\|^{1/2}S^{1/2}_{yy}(A)+\frac{\hat{\Delta}}{1-\hat{\Delta}}\|\hat{S}^{-1}_{xx}(A)\|\|S_{zz}(A)\|^{1/2}S^{1/2}_{yy}(A)\notag\\&\leq\frac{\Delta}{1-\hat{\Delta}}\ub(\|\hat{\beta}(A)\|)+\frac{\hat{\Delta}}{1-\hat{\Delta}}\ub(\|\hat{\beta}(A)\|)\leq\frac{2\hat{\Delta}}{1-\hat{\Delta}}\ub(\|\hat{\beta}(A)\|),
\end{align}
which proves the second part of \eqref{eq:thm-PerturbationOfTheTLSEstimator}.
Note also that the first part of \eqref{eq:thm-PerturbationOfTheTLSEstimator} is a direct consequence of the second part.
This completes the proof.

\section{Proof of Theorem \ref{thm:WTLS-estimator}}
\label{pf:WTLS}
Let $\Delta=\Delta(\hat{\Sigma}^{-1},\Sigma^{-1})$ and $\hat{\Delta}=\ub(\Delta(\hat{S}_{xx}(\hat{\Sigma}^{-1}),\hat{S}_{xx}(\Sigma^{-1})))$.
We only show the second part of \eqref{eq:WTLS-estimator-asymptotic-normality} as the other results in the theorem can be proven similarly.
By \eqref{eq:TLS-decomposition} and Corollary \ref{cor:TLS-estimator-general}, it suffices to show that $\|\hat{\beta}(\hat{\Sigma}^{-1})-\hat{\beta}(\Sigma^{-1})\|=o_{\pr}(p^{-1/2})$.

By a similar argument as that in proving Theorem \ref{thm:TLS-estimator}, $S_{ww}(\Sigma^{-1})\to_{\pr}\mathcal{B}^T Q_0^{**} \mathcal{B} + \Psi$, and thus
\begin{equation}
S_{zz}(\Sigma^{-1})\overset{\pr}{\to}Q^{**}_{0}+\Lambda,\quad\hat{\lambda}(\Sigma^{-1})\overset{\pr}{\to}1,\quad\text{and}\quad S_{yy}(\Sigma^{-1})\overset{\pr}{\to}1+\beta^{T}Q^{**}_{0}\beta.
\end{equation}
This implies that $\hat{S}_{xx}(\Sigma^{-1})\overset{\pr}{\to}Q^{**}_{0}$, and therefore
\begin{equation}
\label{eq:WTLS-estimator-proof-1}
\ub(\|\hat{\beta}(\Sigma^{-1})\|)\overset{\pr}{\to}\frac{\lambda^{1/2}_{\max}(Q^{**}_{0}+\Lambda)(1+\beta^{T}Q^{**}_{0}\beta)^{1/2}}{\lambda_{\min}(Q^{**}_{0})},
\end{equation}
and
\begin{equation}
\label{eq:WTLS-estimator-proof-2}
\frac{\hat{\Delta}}{\Delta}=\frac{\lambda_{\max}(S_{zz}(\Sigma^{-1}))+\hat{\lambda}(\Sigma^{-1})\lambda_{\max}(\Lambda)}{\lambda_{\min}(\hat{S}_{xx}(\Sigma^{-1}))}\overset{\pr}{\to}\frac{\lambda_{\max}(Q^{**}_{0}+\Lambda)+\lambda_{\max}(\Lambda)}{\lambda_{\min}(Q^{**}_{0})}.
\end{equation}
By Condition \ref{asp:trace-WTLS}(i) with $A=\Sigma^{-1}$, $Q_{0}^{**}$ is positive definite, and thus the both limits in \eqref{eq:WTLS-estimator-proof-1} and \eqref{eq:WTLS-estimator-proof-2} are positive constants.
Therefore, it follows from Theorem \ref{thm:PerturbationOfTheTLSEstimator} that
\begin{equation}
\|\hat{\beta}(\hat{\Sigma}^{-1})-\hat{\beta}(\Sigma^{-1})\|\leq\frac{2\hat{\Delta}}{1-\hat{\Delta}}\ub(\|\hat{\beta}(\Sigma^{-1})\|)=O_{\pr}(\Delta)=o_{\pr}(p^{-1/2}),
\end{equation}
as desired.
This completes the proof.

\section{Proof of Theorem \ref{thm:TLS-estimator-nonGaussian}}
\label{pf:TLS-nonGaussian}
We shall adopt the notations used in Appendix \ref{pf:TLS}.
Observe that the asymptotic expansion in \eqref{eq:asymptotic-expression} remains valid under Conditions \ref{asp:moment}(i), \ref{asp:mixing}--\ref{asp:trace}, and \ref{asp:moment-nonGaussian}, as it relies only on the first two moments of $\vecc\mathcal{E}$ and the mixing conditions, which are robust to the non-Gaussianity.
In view of \eqref{eq:asymptotic-expression}, it suffices to reevaluate $\var(S_{we})$ under Condition \ref{asp:moment-nonGaussian} instead of \ref{asp:moment}(ii).
Write
\begin{equation}
\var(S_{we}\mid X)=\var_{\gaus}(S_{we}\mid X)+\var_{\Delta}(S_{we}\mid X),
\label{eq:nonGaussian-decomposition-1}
\end{equation}
where $\var_{\gaus}$ denotes the Gaussian component derived in the proof of Theorem \ref{thm:TLS-estimator}.
It suffices to calculate the non-Gaussian component $\var_{\Delta}$ driven by the parameters $\kappa_{3}$ and $\kappa_{4}$.
Recall that
\begin{equation}
\vecc(S_{we}) =\vecc(\mathcal{B}^T S_{xe} + S_{ee}) = \frac{1}{p}\sum_{i=1}^p \big\{ \mathcal{E}_i \otimes (\mathcal{B}^T X_i) + \mathcal{E}_i \otimes \mathcal{E}_i \big\}.
\end{equation}
By Condition \ref{asp:moment-nonGaussian}, the excess third and fourth moments of $\vecc\mathcal{E}$ over the Gaussian distribution are nonzero only if all respective component indices refer to the same observation $i$ and the same coordinate $j$.
Consequently,
\begin{align}
\var_{\Delta}(p^{1/2}S_{we} \mid X) &= \frac{1}{p}\sum_{i=1}^p \var_{\Delta}(\mathcal{E}_i \otimes \mathcal{E}_i \mid X) \notag \\
&\quad + \frac{1}{p}\sum_{i=1}^p \big\{ \cov_{\Delta}(\mathcal{E}_i \otimes (\mathcal{B}^T X_i), \mathcal{E}_i \otimes \mathcal{E}_i \mid X) + \text{transpose} \big\}.
\label{eq:nonGaussian-decomposition-2}
\end{align}
Note that $\mathcal{E}_i \otimes (\mathcal{B}^T X_i)$ is linear in $\mathcal{E}_i$ and thus the associated variance component depends only on the second moments of $\mathcal{E}_i$, which has no excess effect.

We first derive the kurtosis contribution corresponding to $\var_{\Delta}(\mathcal{E}_i \otimes \mathcal{E}_i \mid X)$.
Observe that the element $E(\mathcal{E}_{ij_{1}}\mathcal{E}_{ij_{2}}\mathcal{E}_{ij_{3}}\mathcal{E}_{ij_{4}}\mid X)$ has an excess effect $E(\mathcal{E}_{ij}^4 \mid X) - 3\{E(\mathcal{E}_{ij}^2 \mid X)\}^2 = \kappa_4 \sigma_i^4 \Psi_{jj}^2$ if and only if $j_{1}=j_{2}=j_{3}=j_{4}=j$.
Let $e_j$ be the $j$-th standard basis vector in $\mathbb{R}^{m+1}$.
Then,
\begin{equation}
\var_{\Delta}(\mathcal{E}_i \otimes \mathcal{E}_i \mid X) = \kappa_4 \sigma_i^4 \sum_{j=1}^{m+1} \Psi_{jj}^2 (e_j \otimes e_j)(e_j \otimes e_j)^T.
\label{eq:excess-kurt-1}
\end{equation}
We project this matrix using $\Pi=(\mathcal{B}^{\perp})^{T}\otimes M$, where $M = Q_0^{-1}(I_m+\Lambda\beta\beta^T)^{-1}\mathcal{L}$.
Indeed,
\begin{equation}
\Pi (e_j \otimes e_j) = \mathcal{B}^\perp_j M e_j.
\label{eq:excess-kurt-2}
\end{equation}
Then, by \eqref{eq:excess-kurt-1} and \eqref{eq:excess-kurt-2}, we have
\begin{equation}
\Pi \var_{\Delta}(\mathcal{E}_i \otimes \mathcal{E}_i \mid X) \Pi^T = \kappa_4 \sigma_i^4 M \Big\{ \sum_{j=1}^{m+1} \Psi_{jj}^2 (\mathcal{B}^\perp_j)^2 e_j e_j^T \Big\} M^T.
\label{eq:excess-kurt-3}
\end{equation}
The matrix within the braces is a diagonal matrix $D_0 = \diag\big( \Psi_{11}^2 (\mathcal{B}^\perp_1)^2, \dots, \Psi_{m+1,m+1}^2 (\mathcal{B}^\perp_{m+1})^2 \big)$.
Observe that for $j \le m$, $\Psi_{jj} = \lambda_j^2$ and $\mathcal{B}^\perp_j = -\beta_j$, yielding the entry $\lambda_j^4 \beta_j^2$, while for $j= m+1$, $\Psi_{m+1,m+1} = 1$ and $\mathcal{B}^\perp_{m+1} = 1$, yielding $1$.
Hence, $D_0 = \diag(D_\beta, 1)$, which implies that
\begin{equation}
\mathcal{L} \Big\{ \sum_{j=1}^{m+1} \Psi_{jj}^2 (\mathcal{B}^\perp_j)^2 e_j e_j^T \Big\} \mathcal{L}^T = (I_m, \Lambda\beta) \begin{pmatrix} D_\beta & 0 \\ 0 & 1 \end{pmatrix} \begin{pmatrix} I_m \\ \beta^T\Lambda \end{pmatrix} = D_\beta + \Lambda\beta\beta^T\Lambda.
\end{equation}
Substituting $M$ back gives
\begin{equation}
M \Big\{ \sum_{j=1}^{m+1} \Psi_{jj}^2 (\mathcal{B}^\perp_j)^2 e_j e_j^T \Big\} M^T = Q_0^{-1}(I_m+\Lambda\beta\beta^T)^{-1} (D_\beta + \Lambda\beta\beta^T\Lambda) (I_m+\beta\beta^T\Lambda)^{-1}Q_0^{-1}.
\label{eq:excess-kurt-4}
\end{equation}
By \eqref{eq:excess-kurt-3}, \eqref{eq:excess-kurt-4}, and Condition \eqref{asp:trace-more}(ii), we have
\begin{equation}
\frac{1}{p}\sum_{i=1}^p \Pi \var_{\Delta}(\mathcal{E}_i \otimes \mathcal{E}_i \mid X) \Pi^T \to \Omega_{\kurt}.
\label{eq:excess-kurt-end}
\end{equation}

Next, we evaluate the skewness contribution corresponding to $\cov_{\Delta}(\mathcal{E}_i \otimes (\mathcal{B}^T X_i), \mathcal{E}_i \otimes \mathcal{E}_i \mid X)$.
Observe that the element $ E(\mathcal{E}_{ij_{1}}(\mathcal{B}^T X_i)_{j_{2}}\mathcal{E}_{ij_3} \mathcal{E}_{ij_4} \mid X)$ has an excess effect $\kappa_3 \sigma_i^3 \Psi_{jj}^{3/2}(\mathcal{B}^T X_i)_{j_{2}}$ if and only if $j_1=j_3=j_4=j$, which implies that
\begin{equation}
\cov_{\Delta}(\mathcal{E}_i \otimes (\mathcal{B}^T X_i), \mathcal{E}_i \otimes \mathcal{E}_i \mid X) = \kappa_3 \sigma_i^3 \sum_{j=1}^{m+1} \Psi_{jj}^{3/2} (e_j \otimes \mathcal{B}^T X_i)(e_j \otimes e_j)^T.
\label{eq:excess-skew-1}
\end{equation}
Projecting the left basis vector $e_a \otimes \mathcal{B}^T X_i$ with $\Pi$ yields
\begin{equation}
\Pi(e_j \otimes \mathcal{B}^T X_i) = \mathcal{B}^\perp_j M \mathcal{B}^T X_i = \mathcal{B}^\perp_j Q_0^{-1}X_i.
\label{eq:excess-skew-2}
\end{equation}
By \eqref{eq:excess-skew-1} and \eqref{eq:excess-skew-2}, we have
\begin{equation}
\Pi E\{\cov_{\Delta}(\mathcal{E}_{i}\otimes(\mathcal{B}^{T}X_{i}),\mathcal{E}_{i}\otimes\mathcal{E}_{i}\mid X)\}\Pi^{T}=\kappa_{3}Q^{-1}_{0}(\sigma^{3}_{i}EX_{i})\Big\{\sum^{m+1}_{j=1}\Psi^{3/2}_{jj}(\mathcal{B}^{\perp}_{j})^{2}e^{T}_{j}\Big\} M^{T}.
\label{eq:excess-skew-3}
\end{equation}
Observe that for $j \le m$, the $j$-th summand is $\lambda_j^3 \beta_j^2 e_j^T$, while for $j = m+1$, the summand is $e_{m+1}^T$.
Hence, the summation results in $(\gamma_\beta^T, 1)$, which implies that
\begin{equation}
\Big\{ \sum_{j=1}^{m+1} \Psi_{jj}^{3/2} (\mathcal{B}^\perp_j)^2 e_j^T \Big\}M^T = (\gamma_\beta^T, 1) \begin{pmatrix} I_m \\ \beta^T\Lambda \end{pmatrix}(I_m+\beta\beta^T\Lambda)^{-1}Q_0^{-1}=\nu_\beta^TQ_0^{-1}.
\label{eq:excess-skew-4}
\end{equation}
By \eqref{eq:excess-skew-3}--\eqref{eq:excess-skew-4} and Condition \cref{asp:trace-more}(i), we have
\begin{equation}
\frac{1}{p}\sum^{p}_{i=1}\Pi E\{\cov_{\Delta}(\mathcal{E}_{i}\otimes(\mathcal{B}^{T}X_{i}),\mathcal{E}_{i}\otimes\mathcal{E}_{i}\mid X)\}\Pi^{T}\to\kappa_{3}Q^{-1}_{0}\mu_{x}\nu^{T}_{\beta}Q^{-1}_{0}.
\label{eq:excess-skew-end}
\end{equation}
Adding the transpose from the symmetric part gives $\kappa_3 Q_0^{-1} (\mu_x \nu_\beta^T + \nu_\beta \mu_x^T) Q_0^{-1} = \Omega_{\skeww}$.
Then, by combining \eqref{eq:asymptotic-expression}, \eqref{eq:nonGaussian-decomposition-1}, \eqref{eq:nonGaussian-decomposition-2}, \eqref{eq:excess-kurt-end}, and \eqref{eq:excess-skew-end}, we complete the proof.

\section{Proof of Theorem \ref{thm:HAC-estimator}}
\label{pf:HAC}
We shall adopt the notations used in Appendix \ref{pf:TLS}.
As $S_{xx}\to_{\pr}Q_{0}$ and $\hat{\beta}\to_{\pr}\beta$ by Theorem \ref{thm:TLS-estimator-nonGaussian-general}, it suffices to show that $\hat{V}_{g}\to_{\pr}V_{g}$.
Observe that by Lemma \ref{lem:KvR05-Kronecker},
\begin{equation}
g_{i}=\mathcal{L}W_{i}W_{i}^{T}\mathcal{B}^{\perp}=\{(\mathcal{B}^{\perp})^{T}\otimes\mathcal{L}\}\vecc(W_{i}W_{i}^{T}).
\end{equation}
Let $\hat{\mathcal{L}}=(I_{m},\Lambda\hat\beta)$ and $\hat{\mathcal{B}}^{\perp}=(-\hat\beta^T,1)^T$.
By a similar argument, we have
\begin{align}
\hat{V}_{g}&=\{(\hat{\mathcal{B}}^{\perp})^{T}\otimes\hat{\mathcal{L}}\}\Big\{\frac{1}{p}\sum_{i=1}^p \sum_{j=1}^p\vecc(W_{i}W_{i}^{T})\vecc^T(W_{j}W_{j}^{T})\mathcal{K}\Big(\frac{\| s_{i}-s_{j}\|}{h_{p}}\Big)\Big\}(\hat{\mathcal{B}}^{\perp}\otimes\hat{\mathcal{L}}^T)\notag\\&=\{({\mathcal{B}}^{\perp})^{T}\otimes{\mathcal{L}}\}\Big\{\frac{1}{p}\sum_{i=1}^p \sum_{j=1}^p\vecc(W_{i}W_{i}^{T})\vecc^T(W_{j}W_{j}^{T})\mathcal{K}\Big(\frac{\| s_{i}-s_{j}\|}{h_{p}}\Big)\Big\}({\mathcal{B}}^{\perp}\otimes{\mathcal{L}}^T)\{1+o_{\pr}(1)\}\notag\\&=\Big\{\frac{1}{p}\sum_{i=1}^p \sum_{j=1}^p g_i g_j^T \mathcal{K}\Big(\frac{\| s_{i}-s_{j}\|}{h_{p}}\Big)\Big\}\{1+o_{\pr}(1)\}=:\tilde{V}_{g}\{1+o_{\pr}(1)\}.
\end{align}
Then, it suffices to show that $\tilde{V}_{g}\to_{\pr}V_{g}$.
Let $\check{V}_{g}=p^{-1}\sum_{i=1}^p \sum_{j=1}^p g_i g_j^T=pS_gS_g^T$.
As $ES_g=0$ by \eqref{eq:Sww-zero-mean}, we have $E\check{V}_g=\var(p^{1/2}S_g)\to V_g$ by Condition \ref{asp:moment-nonGaussian-general}(ii).
Then, it suffices to show that
\begin{equation}
\label{eq:Vg-1}
\tilde{V}_{g}-E\tilde{V}_{g}\overset{\pr}{\to}0,
\end{equation}
and that
\begin{equation}
\label{eq:Vg-2}
E\tilde{V}_{g}-E\check{V}_{g}\overset{\pr}{\to}0.
\end{equation}

We first prove \eqref{eq:Vg-2}, which requires the following three additional lemmas taken from \cite{Bra07-MR2325294} and \cite{JP09-MR2525996}.
Let $Q_{X}(u)=\inf\{x:\pr(X\leq x)\geq1-u\}$ be the upper quantile function of $X$.
Note that $Ef(X)=\int_0^1f(Q_X(u))du$ for any Borel $f$ with $E|f(X)|<\infty$.
Let $a\lesssim b$ denote $a\leq Cb$ for a constant $C>0$.

\begin{lemma}[Lemma A.1(iii) in \citealp{JP09-MR2525996}]
\label{lem:JP09-counting}
Let $D\subset\mathbb{R}^d$ satisfy Condition \ref{asp:mixing}(i) with $\rho_0>0$. For $k\ge0$, let $N_s(k)=\{t\in D: k \le \|s - t\| < k+1\}$ be the collection of the $k$-th layer neighbors of $s\in D$. Then, there is a constant $C>0$ depending only on $\rho_0$ and $d$ such that $|N_s(k)| \le Ck^{d-1}$ for every $k\ge1$ and $s\in D$.
\end{lemma}

\begin{lemma}[{Rio's inequality; \citealp[page 320]{Bra07-MR2325294}}]
\label{lem:rio-ineq}
Suppose that $X$ and $Y$ are two random variables such that $E|X|<\infty$ and $E|Y|<\infty$, and that $\int_0^1Q_{|X|}(u)Q_{|Y|}(u)du<\infty$.
Let $\alpha=\alpha(\sigma(X),\sigma(Y))$ be the $\alpha$-mixing coefficient between the $\sigma$-algebras generated by $X$ and $Y$.
Then, we have $|\cov(X,Y)|\leq4\int_0^\alpha Q_{|X|}(u)Q_{|Y|}(u)du$.
\end{lemma}

\begin{lemma}[Lemma B.1 in \citealp{JP09-MR2525996}]
\label{lem:JP09-integral}
Let $\{\alpha(k): k\ge1\}$ be a nonincreasing sequence such that $0 \le \alpha(k) \le 1$ and $\alpha(k) \to 0$ as $k\to\infty$. Set $\alpha(0)=1$ and $\alpha_{\inv }(u) = \max\{k\ge0: \alpha(k) > u\}$ for $0 < u < 1$. Let $f: (0, 1) \to [0, \infty)$ be a Borel function. Then, for every $q\ge1$, we have $\sum_{k=1}^\infty k^{q-1}\int_0^{\alpha(k)} f(u) du \le \int_0^1 \alpha_{\inv }^q(u) f(u) du$ and $\int_0^1 \alpha_{\inv }^q(u) du \le q\sum_{k=1}^\infty k^{q-1}\alpha(k)$.
\end{lemma}

Returning to the proof of \eqref{eq:Vg-2}, we can write the difference as
\begin{equation}
E\tilde{V}_g - E\check{V}_g = \frac{1}{p}\sum_{i=1}^p \sum_{j=1}^p E(g_i g_j^T) \Big\{ \mathcal{K}\Big(\frac{\|s_i - s_j\|}{h_p}\Big) - 1 \Big\}.
\end{equation}
By Condition \ref{asp:kernel}(ii), the collection $\{\|g_i\|^{4+2\delta}: i \ge 1\}$ is uniformly integrable.
By Lemma \ref{lem:rio-ineq},
\begin{equation}
\|E(g_i g_j^T)\| \lesssim \int_{0}^{\alpha(\|s_i - s_j\|)} Q_{\|g_i\|}(u) Q_{\|g_j\|}(u) du \lesssim \int_{0}^{\alpha(\|s_i - s_j\|)} \big\{ Q_{\|g_i\|}^2(u) + Q_{\|g_j\|}^2(u) \big\} du.
\end{equation}
Summing over $j$, we notice that by definition the condition $u < \alpha(\|s_i - s_j\|)$ is equivalent to $\|s_i - s_j\| \le \alpha_{\inv }(u)$. By Lemma \ref{lem:JP09-counting}, the number of such $j$ is bounded by $\sum_{k=0}^{\lfloor \alpha_{\inv }(u) \rfloor} C\{k^{d-1} + I(k=0)\} \lesssim \alpha_{\inv }^d(u) + 1$.
Consequently,
\begin{equation}
\sum_{j=1}^p \int_0^{\alpha(\|s_i - s_j\|)} Q_{\|g_i\|}^2(u) du \lesssim \int_0^1 (\alpha_{\inv }^d(u) + 1) Q_{\|g_i\|}^2(u) du.
\end{equation}
By H\"older's inequality, Lemma \ref{lem:JP09-integral}, and Condition \ref{asp:mixing}(ii), we have
\begin{align}
\int_0^1 \alpha_{\inv }^d(u) Q_{\|g_i\|}^2(u) du &\le \Big( \int_0^1 \alpha_{\inv }^{(1+2/\delta)d}(u) du \Big)^{\delta/(2+\delta)} \Big( \int_0^1 Q_{\|g_i\|}^{2+\delta}(u) du \Big)^{2/({2+\delta})}\notag\\
&\lesssim\Big(\sum_{k=1}^{\infty}k^{(1+2/\delta)d-1}\alpha(k)\Big)^{\delta/(2+\delta)}(E\|g_i\|^{2+\delta})^{2/(2+\delta)}<\infty.
\end{align}
By symmetry, an identical argument applies to the summation involving $Q_{\|g_j\|}^2(u)$. Therefore, we can conclude that $\sum_{i=1}^{p}\sum_{j=1}^{p}\|E(g_ig_j^T)\|\lesssim p$.

Let $K>0$. Observe that under $u < \alpha(\|s_i-s_j\|)$, we have $\|s_{i}-s_{j}\|>K$ implies that $u < \alpha(K)$, or equivalently $\alpha_{\inv }(u) > K$.
Then, by a similar argument,
\begin{align}
&\frac{1}{p}\sum^{p}_{i=1}\sum^{p}_{j=1}\|E(g_{i}g^{T}_{j})\|I(\|s_{i}-s_{j}\|>K)\notag\\&\lesssim\frac{1}{p}\sum^{p}_{i=1}\sum^{p}_{j=1}\int^{\alpha(\|s_{i}-s_{j}\|)}_{0}Q^{2}_{\|g_{i}\|}(u)I(\alpha_{\inv}(u)>K)du\notag\\&\lesssim\frac{1}{p}\sum^{p}_{i=1}\int^{1}_{0}I(\alpha_{\inv}(u)>K)(\alpha^{d}_{\inv}(u)+1)Q^{2}_{\|g_{i}\|}(u)du\notag\\&\leq\Big\{\int^{1}_{0}I(\alpha_{\inv}(u)>K)(\alpha^{d}_{\inv}(u)+1)^{1+2/\delta}du\Big\}^{\delta/(2+\delta)}\sup_{i\geq1}(E\|g_{i}\|^{2+\delta})^{2/(2+\delta)}\notag\\&\lesssim\Big\{\int^{1}_{0}I(\alpha_{\inv}(u)>K)(\alpha^{d}_{\inv}(u)+1)^{1+2/\delta}du\Big\}^{\delta/(2+\delta)},
\end{align}
which converges to $0$ as $K\to\infty$ by the dominated convergence theorem.
Consequently,
\begin{align}
\|E\tilde{V}_{g}-E\check{V}_{g}\|&\le\frac{1}{p}\Big\{\sum_{i,j:\|s_{i}-s_{j}\|\leq K}+\sum_{i,j:\|s_{i}-s_{j}\|>K}\Big\}\|E(g_{i}g^{T}_{j})\|\Big|\mathcal{K}\Big(\frac{\|s_{i}-s_{j}\|}{h_{p}}\Big)-1\Big|\notag\\&\lesssim\sup_{x\le K/h_{p}}|\mathcal{K}(x)-1|+\frac{1}{p}\sum_{i,j:\|s_{i}-s_{j}\|>K}\|E(g_{i}g^{T}_{j})\|.
\end{align}
By Condition \ref{asp:kernel}(i), the first term converges to $0$ as $h_p\to\infty$ for each $K>0$.
Then, \eqref{eq:Vg-2} follows from taking $p\to\infty$ first, and then taking $K\to\infty$.

Next, it suffices to establish \eqref{eq:Vg-1}. Let $Y_{ij} = g_{ia}g_{ib}\mathcal{K}(\|s_i - s_j\| / h_p)$ for any fixed matrix indices $a, b \in \{1, \dots, m+1\}$. We have $\tilde{V}_{g,ab} = p^{-1} \sum_{i}^p\sum_{j}^p Y_{ij}$.
Let $r_{ijlm} = \rho(\{s_i, s_j\}, \{s_l, s_m\}) = \min_{u \in \{i,j\}, v \in \{l,m\}} \|s_u - s_v\|$. By Lemma \ref{lem:rio-ineq},
\begin{equation}
|\cov(Y_{ij},Y_{lm})|\le4\int^{\alpha(r_{ijlm})}_{0}Q_{|Y_{ij}|}(u)Q_{|Y_{lm}|}(u)du\le2\int^{\alpha(r_{ijlm})}_{0}\big\{ Q^{2}_{|Y_{ij}|}(u)+Q^{2}_{|Y_{lm}|}(u)\big\} du.
\end{equation}
By symmetry, summing over $i, j, l, m$ yields:
\begin{equation}
|\var(\tilde{V}_{g,ab})|\leq\frac{1}{p^{2}}\sum^{p}_{i,j,l,m=1}|\cov(Y_{ij},Y_{lm})|\leq\frac{4}{p^{2}}\sum^{p}_{i,j,l,m=1}\int^{\alpha(r_{ijlm})}_{0}Q^{2}_{|Y_{ij}|}(u)du.
\end{equation}
For a fixed pair $(i, j)$ where $Y_{ij} \neq 0$ (implying $\|s_i - s_j\| \le h_p$), we first count the number of pairs $(l, m)$ satisfying $\lfloor r_{ijlm} \rfloor = k$ and $Y_{lm} \neq 0$ (implying $\|s_l - s_m\| \le h_p$).
If $\lfloor r_{ijlm} \rfloor = k$, at least one of $s_l$ and $s_m$ is at a distance within $[k, k+1)$ from $s_i$ or $s_j$. By Lemma \ref{lem:JP09-counting}, the number of choices for this first site (say, $s_l$) is bounded by $C \max\{1, k^{d-1}\}$. Once $l$ is fixed, the restriction $\|s_l - s_m\| \le h_p$ leaves at most $C h_p^d$ choices for $m$. Then, the total number of such pairs $(l, m)$ is bounded by $C h_p^d \max\{1, k^{d-1}\}$.
Consequently, by Lemma \ref{lem:JP09-integral} and H\"older's inequality,
\begin{align}
&\sum^{p}_{l=1}\sum^{p}_{m=1}\int^{\alpha(r_{ijlm})}_{0}Q^{2}_{|Y_{ij}|}(u)du\notag\\&\lesssim\sum^{\infty}_{k=0}h^{d}_{p}\max\{1,k^{d-1}\}\int^{\alpha(k)}_{0}Q^{2}_{|Y_{ij}|}(u)du\notag\\&\lesssim h^{d}_{p}\int^{1}_{0}(\alpha^{d}_{\inv}(u)+1)Q^{2}_{|Y_{ij}|}(u)du\notag\\&\lesssim h^{d}_{p}\Big\{\int^{1}_{0}(\alpha^{d}_{\inv}(u)+1)^{1+2/\delta}du\Big\}^{\delta/(2+\delta)}\sup_{i,j\geq1}\{E|Y_{ij}|^{2+\delta}\}^{2/(2+\delta)}\notag\\&\lesssim h^{d}_{p}\Big(\sum^{\infty}_{k=1}k^{(1+2/\delta)d-1}\alpha(k)\Big)^{\delta/(2+\delta)}\sup_{i\geq1}(E\|g_{i}\|^{4+2\delta})^{2/(2+\delta)}\lesssim h^{d}_{p},
\end{align}
where the last step follows from Condition \ref{asp:mixing}(ii) and the uniform integrability of $\{\|g_i\|^{4+2\delta}\}$.

Finally, we take the summation over $i$ and $j$. Since $Y_{ij}$ is nonzero only if $\|s_i - s_j\| \le h_p$, there are at most $Ch_p^d$ valid choices of $j$ for each of the $p$ choices of $i$. The total number of nonzero pairs $(i, j)$ is therefore at most $Cp h_p^d$.
Therefore, we have $|\var(\tilde{V}_{g,ab})|\lesssim h^{2d}_{p}/p\to0$ when $h_p=o(p^{1/2d})$, as desired.
This completes the proof.
%
%

%
%

\end{document}